\documentclass[a4paper]{article}

\usepackage[utf8]{inputenc}
\usepackage{graphicx}
\usepackage{tikz}
\usepackage{amssymb}
\usepackage{amsmath}
\usepackage{bbm}
\usepackage[nameinlink, capitalise, noabbrev]{cleveref}
\usepackage{theorem}
\newtheorem{definition}{Definition}
\newtheorem{example}{Example}

\let\plainitem=\item
\let\plainitemitem=\itemitem
\def\today{\number\day.\number\month.\number\year}
\def\openface{\Bbb}                
\def\N{{\openface N}}              

\def\R{{\openface R}}
\def\C{{\openface C}}
\def\g{\hskip.17em\relax}               
\def\th{\thinspace}                     
\def\nl{\hfil\break}
\newskip\Bigskipamount
\def\Bigbreak{\removelastskip\vskip0pt plus .1\vsize\penalty-1000
              \vskip0pt plus-.1\vsize\vskip\Bigskipamount}
\def\Nobreak$$#1$${\postdisplaypenalty=10000$$#1$$\postdisplaypenalty=0}
\let\doublebar=\| 
\def\|{\!\!\restriction\!\!}

  \let\sub=\sube

\def\supe{\supseteq}

\def\sm{\smallsetminus}
\def\es{\emptyset}

\def\T{T} \def\H{I}
\def\wrt{with respect to}

\def\:{\colon}
\def\minor{\preccurlyeq} 
\def\Minor{\succcurlyeq}

\def\slt{\mathrel{\hbox{$\minor$\kern-.6em\lower.33ex\hbox{${}_s\;$}}}}
\def\sgt{\mathrel{\mathchoice                        
   {\hbox{$\Minor$\kern-.5em\lower.3ex\hbox{${}_s$}}}
   {\hbox{$\Minor$\kern-.5em\lower.3ex\hbox{${}_s$}}}
   {\hbox{$\scriptstyle\Minor\kern-.43em\lower.28ex\hbox{$\scriptstyle{}_s$}$}}
 {\hbox{$\scriptstyle\Minor\kern-.43em\lower.28ex\hbox{$\scriptstyle{}_s$}$}} }}    

\def\ucl(#1){\lfloor #1 \rfloor}
\def\dcl(#1){\lceil #1 \rceil}
\def\specrel#1#2{\mathrel{\mathop{\kern0pt #1}\limits_{#2}}}
\def\Specrel#1#2{\mathrel{\mathop{\kern0pt #1}\limits^{#2}}}
\def\alignspecrel#1#2{\mathrel{\mathop{\kern0pt #1}\limits_{\hbox
   to0pt{\hss$\scriptstyle#2$\hss}}}}
\def\alignSpecrel#1#2{\mathrel{\mathop{\kern0pt #1}\limits^{\hbox
   to0pt{\hss$\scriptstyle#2$\hss}}}}
\def\invlim{\specrel\lim{\raise 2pt\hbox{$\longleftarrow$}}}
\def\proclaimwithname #1. (#2) #3\par{{\bigbreak
  \clubpenalty=10000\noindent{\bf#1.\enspace}(#2)\nl
  {\sl #3}\par\bigbreak}}
\def\proposition (#1) #2\par{{\setbox0\hbox{(#1)\enspace}\bigbreak
   \sl\hangindent\the\wd0 \noindent\hskip\the\wd0
   \llap{\box0}\ignorespaces#2\par\bigbreak}}
\def\beginpsection #1\par{\Bigbreak\centerline{\bold #1}
        \penalty10000\bigskip}
\def\psubsection #1\par{\bigbreak\leftline{\bold #1}\penalty10000\bigskip}
\def\pitem#1{\smallskip\advance\parindent by 3mm
             \plainitem{\rm(#1)}\advance\parindent by-3mm}
\def\pitemitem#1{\smallskip\advance\parindent by 3mm
             \plainitemitem{\rm(#1)}\advance\parindent by-3mm}
\def\varitemitem#1{{\setbox0\hbox{\hskip\parindent#1\enskip}
           \smallbreak\hangindent\the\wd0 \noindent\hskip\the\wd0
           \llap{#1\enskip}\ignorespaces}}
\newdimen\newparindent
\def\iitem#1#2\par{\newparindent=\parindent \advance\newparindent by 3mm
           \smallbreak \hangindent\newparindent \noindent\hskip\newparindent
           \llap{{\rm #1}\enspace}\ignorespaces#2\par\smallbreak}
\def\iitemitem#1#2\par{\newparindent=\parindent \advance\newparindent by 3mm
           \smallbreak \hangindent2\newparindent \noindent\hskip2\newparindent
           \llap{{\rm #1}\enspace}\ignorespaces#2\par\smallbreak}
\def\varitem#1#2\par{{\setbox0\hbox{{\rm #1}\enspace}
           \smallbreak \hangindent\the\wd0 \noindent\hskip\the\wd0
           \llap{{\rm #1}\enspace}\ignorespaces#2\par\smallbreak}}

\def\Textindent#1{\par \advance\parindent by 3mm
                  \textindent{{\rm #1}} \advance\parindent by -3mm}
\def\indentedline#1{\advance\hsize by -\parindent \line{#1}
                   \advance\hsize by \parindent}
\def\iindentedline#1{\advance\parindent by 3mm
                     \advance\hsize by -\parindent
                     \line{#1}
                     \advance\hsize by \parindent
                     \advance\parindent by -3mm}
\newdimen\margin   
\def\textdisplay#1&#2&#3$${\margin=\hsize
          \setbox1=\hbox{$\displaystyle#1\quad$}%
          \setbox2=\hbox{\quad#2\qquad$#3$}%
                     \advance\margin by-\wd1
                     \divide\margin by 2
   \ifdim\wd2 < \margin
      \box1\rlap{\quad#2}\eqno#3$$%
   \else
      \line{\qquad\hfil \box1\quad #2 \qquad $#3$}$$%
   \fi}
\def\ltextdisplay#1&#2&#3$${\margin=\hsize
           \setbox2=\hbox{$\displaystyle#2\quad$}
           \setbox3=\hbox{\quad#3\qquad}
                     \advance\margin by-\wd2
                     \divide\margin by 2
   \ifdim\wd3 < \margin
      \line{$#1$\hfil\box2\hbox to \margin{\box3\hfil}}$$%
   \else
      \line{$#1$\qquad\hfil\box2\quad #3\qquad} $$%
   \fi}
\def\textno#1&#2\par{%
   \margin=\hsize
   \advance\margin by -4\parindent
          \setbox1=\hbox{\sl#1}%
   \ifdim\wd1 < \margin
      $$\box1\eqno#2$$\endgraf%
   \else
      \bigbreak
      \line{\indent$\vcenter{\advance\hsize by -3\parindent
      \sl\noindent#1}\hfil#2$}%
      \bigbreak
   \fi}
\def\textlno#1&#2\par{%
   \margin=\hsize
   \advance\margin by -4\parindent
          \setbox1=\hbox{\sl#1}%
   \ifdim\wd1 < \margin
      $$\box1\leqno#2$$%
   \else
      \bigbreak
      \line{$#2\hfil\vcenter{\advance\hsize by -3\parindent
          \sl\noindent#1}\hskip\parindent$}%
      \bigbreak
   \fi}
\def\textnoNonPar#1&#2\endgraf{%
   \margin=\hsize
   \advance\margin by -4\parindent
          \setbox1=\hbox{\sl#1}%
   \ifdim\wd1 < \margin
      $$\box1\eqno#2$$\endgÃ¯Â¿Â½raf%
   \else
      \bigbreak
      \line{\indent$\vcenter{\advance\hsize by -3\parindent
      \sl\noindent#1}\hfil#2$}%
      \bigbreak
   \fi}
\newcount\commentno
\def\COMMENT#1{$^{<\the\commentno>}$%
     \vadjust{\vbox to 0pt{\vss\vskip-8pt\rightline{%
     \rlap{\hbox{\hskip7mm \vbox{\pretolerance=-1
     \doublehyphendemerits=0 \finalhyphendemerits=0
     \hsize40mm\tolerance=10000\eightpoint
     \lineskip=0pt\lineskiplimit=0pt
     \rightskip=0pt plus16mm\baselineskip8pt\noindent
     \hskip0pt       
     {$\langle$\the\commentno. #1$\rangle$}\endgraf}}}}\vss}}%
     \global\advance\commentno by1}%
\def\writecommentsasfootnotes{%
 \def\COMMENT{\global\advance\commentno by1\footnote{$^{<\the\commentno>}$}}%
 }
\def\nocomments{\def\COMMENT##1{}}
\def\?#1{\vadjust{\vbox to 0pt{\vss\vskip-8pt\leftline{%
     \llap{\hbox{\vbox{\pretolerance=-1
     \doublehyphendemerits=0\finalhyphendemerits=0
     \hsize30truemm\tolerance=10000\small
     \lineskip=0pt\lineskiplimit=0pt
     \rightskip=0pt plus16truemm\baselineskip8pt\noindent
     \hskip0pt        
     #1\endgraf}\hskip7truemm}}}\vss}}}
\def\d{}
\def\ds#1{}
\long\def\indexwrite#1{%
    \thingtowrite={#1}%
    \immediate\write\index{\the\thingtowrite}%
    }
\newwrite\index
\def\makeindex{\immediate\openout\index=index%
   \immediate\write\index{\catcode`@=11}%
   \def\d##1 {\ifmmode
     \write\index{$##1$, }%
     \write\index{\the\count0}\write\index{}
   \else
     \write\index{{##1}, }%
     \write\index{\the\count0}\write\index{}
   \fi {##1} }
      \def\ds##1{\ifmmode
     \write\index{$##1$, }%
     \write\index{\the\count0}\write\index{}
   \else
     \write\index{##1, }%
     \write\index{\the\count0}\write\index{}
   \fi}}
\newdimen\gap
\newdimen\hackwidth
\def\mo#1{\ifmmode {#1}\else {\it#1}\fi\mos{#1}}
\def\mos#1{\ifmmode
     \strut\vadjust{\vbox to 0pt{\vss\kern-11pt\leftline{%
     \llap{\hbox{\vbox{\pretolerance=-1
     \doublehyphendemerits=0\finalhyphendemerits=0
     \hsize\hackwidth\tolerance=10000\eightpoint
     \lineskip=0pt\lineskiplimit=0pt
     \rightskip=0pt plus\hsize\baselineskip8pt\noindent
     $#1$\strut\endgraf}\hskip\gap }}}\vss}}%
   \else
     \strut\vadjust{\vbox to 0pt{\vss\kern-11pt\leftline{%
     \llap{\hbox{\vbox{\pretolerance=-1
     \doublehyphendemerits=0\finalhyphendemerits=0
     \hsize\hackwidth\tolerance=10000\eightpoint
     \lineskip=0pt\lineskiplimit=0pt
     \rightskip=0pt plus\hsize\baselineskip8pt\noindent
     \hskip0pt    
     {\sl#1}\strut\endgraf}\hskip\gap }}}\vss}}%
   \fi}%
\newcount\remarkno
\def\REMARK#1{{\footnote{${}^{\the\remarkno}$}{{#1}}%
   \global\advance\remarkno by1}}

\newif\ifTextures
\newdimen\topfiguremargin
\newdimen\bottomfiguremargin
   \bottomfiguremargin=\medskipamount                    
\newdimen\normalpictureheight
\def\Fig.#1 (#2by#3; heightfactor:#4; caption:#5) {{%
   \dimen2=\normalpictureheight
   \dimen0=#2                          
      \divide\dimen2 by 1000
      \multiply\dimen2 by#4              
   \count2=\dimen2                  
      \dimen1=#3                             
   \count1=\dimen1
   \divide\count1 by 1000
   \divide\count2 by \count1          
   \divide\dimen0 by 1000
   \multiply\dimen0 by \count2      
         \dimen1=\hsize
         \advance\dimen1 by -\dimen0
         \divide\dimen1 by 2               
   \midinsert
   \vbox to \topfiguremargin{\vfil}
   \noindent\hskip\dimen1
   \picture\dimen0 by \dimen2  (Fig.#1 scaled \the\count2)%
   \vskip\bottomfiguremargin                     
      \ninepoint
      \parindent=.1\hsize\narrower\narrower
      \setbox0\hbox{#5}
      \ifdim\wd0 < .6\hsize
           \centerline{F{\sc IGURE} #1.\hskip1em#5}
       \else
           \plainitem{F{\sc IGURE} #1. }#5\par
       \fi
   \vskip0pt\endinsert}}
\def\textpicture #1(#2by#3; #4width#5lower#6){{%
      \dimen0=#5\count2=\dimen0                    
      \dimen0=#2\count1=\dimen0                    
   \divide\count1 by 1000
   \divide\count2 by \count1                 
   \hbox{\vrule #4width0pt\vbox to 0pt{\vss\vskip#6%
      \special{picture #1 scaled \the\count2}\hrule width#5 height0pt\vss}}}}
\def\Figure #1. #2 (#3; #4) {{
   \def\bigskip{\par\ifdim\lastskip<\bigskipamount\removelastskip
      \vskip\bigskipamount\fi}
   \midinsert\vskip\topfiguremargin
   \dimen0=\normalpictureheight
      \divide\dimen0 by 1000
      \multiply\dimen0 by#4        
   \centerline{\includegraphics[scale=1]{#3.pdf}}
   \vskip\bottomfiguremargin                     
      \ninepoint
      \parindent=.1\hsize\narrower\narrower
      \setbox0\hbox{#2}
      \ifdim\wd0 < .6\hsize
           \centerline{F{\sc IGURE} #1.\hskip1em#2}
       \else
           \plainitem{F{\sc IGURE} #1. }#2\par
       \fi
  \endinsert}}
\def\Abh#1 {{\sl Abh.\g Math.\g Sem.\g Univ.\g Hamburg\penalty100\ \bf#1\ }}
\def\AMASH#1 {{\sl Acta Math.\g Acad.\g Sci.\g Hung.\penalty100\ \bf#1\ }}
\def\Advances#1 {{\sl Adv.\g Math.\penalty100\ \bf#1\ }}
\def\Annals#1 {{\sl Ann.\g Math.\penalty100\ \bf#1\ }}
\def\ActaArith#1 {{\sl Acta\g Arith.\penalty100\ \bf#1\ }}
\def\AnnComb#1 {{\sl Ann.\g Comb.\penalty100\ \bf#1\ }}
\def\AMM#1 {{\sl Amer.\g Math.\g Monthly\penalty100\ \bf#1\ }}
\def\Archiv#1 {{\sl Arch.}\g {\sl Math.\penalty100\ \bf#1\ }}
\def\ArsComb#1 {{\sl Ars Comb.\penalty100\ \bf#1\ }}
\def\CJM#1 {{\sl Can.\g J.\th Math.\penalty100\ \bf#1\ }}
\def\Comb#1 {{\sl Com\-bi\-na\-to\-ri\-ca\penalty100\ \bf#1\ }}
\def\CPC#1 {{\sl Comb.\g Probab.\g Comput.\penalty100\ \bf#1\ }}
\def\Crelle#1 {{\sl J.}\th {\sl Reine Angew.}\g
    {\sl Math.\penalty100\ \bf#1\ }}
\def\DM#1 {{\sl Discrete Math.\penalty100\ \bf#1\ }}
\def\DAM#1 {{\sl Discrete Appl.\g Math.\penalty100\ \bf#1\ }}
\def\EJC#1 {{\sl Eur.}\g{\sl J.}\g{\sl Comb.\penalty100\ \bf#1\ }}
\def\EJ#1 {{\sl Electronic.}\g{\sl J.}\g{\sl Comb.\penalty100\ \bf#1\ }}
\def\GC#1 {{\sl Graphs Comb.\penalty100\ \bf#1\ }}
\def\IJ#1 {{\sl Isr.\g J.\th Math.\penalty100\ \bf#1\ }}
\def\Inv#1 {{\sl Inv.\g math.\penalty100\ \bf#1\ }}
\def\JAlg#1 {{\sl J.}\th {\sl Algorithms\penalty100\ \bf#1\ }}
\def\JCTA#1 {{\sl J.}\th {\sl Comb.}\g {\sl Theory~A\penalty100\ \bf#1\ }}
\def\JCTB#1 {{\sl J.}\th {\sl Comb.}\g {\sl Theory~B\penalty100\ \bf#1\ }}
\def\JOC#1 {{\sl J.}\th {\sl Combinatorics\penalty100\ \bf#1\ }}
\def\JGT#1 {{\sl J.}\th {\sl Graph Theory\penalty100\ \bf#1\ }}
\def\BLMS#1 {{\sl Bull.\g Lond.\g Math.\g Soc.\penalty100\ \bf#1\ }}
\def\JLMS#1 {{\sl J.\g Lond.\g Math.\g Soc.\penalty100\ \bf#1\ }}
\def\PLMS#1 {{\sl Proc.\g Lond.\g Math.\g Soc.\penalty100\ \bf#1\ }}
\def\Order#1 {{\sl Order\ \bf#1\ }}
\def\Random#1 {{\sl Random Struct.\g Alg.\penalty100\ \bf#1\ }}
\def\MA#1 {{\sl Math.}\g {\sl Ann.\penalty100\ \bf#1\ }}
\def\MN#1 {{\sl Math.}\g {\sl Nachr.\penalty100\ \bf#1\ }}
\def\MPCPS#1 {{\sl Math.\g Proc.\g Camb.\g Phil.\g Soc.\penalty100\ \bf#1\ }}
\def\MS#1 {{\sl Math.}\g {\sl Scand.\penalty100\ \bf#1\ }}
\def\MZ#1 {{\sl Math.}\g {\sl Zeit.\penalty100\ \bf#1\ }}
\def\BAMS#1 {{\sl Bull.\th Amer.\g Math.\g Soc.\penalty100\ \bf#1\ }}
\def\JAMS#1 {{\sl J.\th Amer.\g Math.\g Soc.\penalty100\ \bf#1\ }}
\def\MAMS#1 {{\sl Mem.\g Amer.\g Math.\g Soc.\penalty100\ \bf#1\ }}
\def\PAMS#1 {{\sl Proc.\g Amer.\g Math.\g Soc.\penalty100\ \bf#1\ }}
\def\SIAM#1 {{\sl SIAM J.\g Discrete Math.\penalty100\ \bf#1\ }}
\def\SLNM#1 {{\sl Springer Lecture Notes in Mathematics\penalty100\ \bf#1\ }}
\def\TAMS#1 {{\sl Trans.\g Amer.\g Math.\g Soc.\penalty100\ \bf#1\ }}
\def\TCSA#1 {{\sl Theor.\g Comput.\g Sci.~A\penalty100\ \bf#1\ }}
\nocomments

\def\T{\mathcal T}
\def\H{\mathbb{H}}
\def\R{\mathbb{R}}
\def\I{\mathbb{I}}
\def\lowfwd #1#2#3{{\mathop{\kern0pt #1}\limits^{\kern#2pt\raise.#3ex
\vbox to 0pt{\hbox{$\scriptscriptstyle\rightarrow$}\vss}}}}
\def\lowbkwd #1#2#3{{\mathop{\kern0pt #1}\limits^{\kern#2pt\raise.#3ex
\vbox to 0pt{\hbox{$\scriptscriptstyle\leftarrow$}\vss}}}}
\def\fwd #1#2{{\lowfwd{#1}{#2}{15}}}
\def\vSk{\lowfwd {S_k}11}

\def\vS{{\hskip-1pt{\fwd S3}\hskip-1pt}}

\def\ve{\kern-1.5pt\lowfwd e{1.5}2\kern-1pt}
\def\ev{\kern-1pt\lowbkwd e{0.5}2\kern-1pt}
\def\vf{\kern-2pt\lowfwd f{2.5}2\kern-1pt}

\def\vr{\lowfwd r{1.5}2}

\def\vs{\lowfwd s{1.5}1}
\def\sv{\lowbkwd s{0}1}
\def\vsk{\lowfwd {s_k}11}
\def\svk{\lowbkwd {s_k}12}
\def\vSd{{\mathop{\kern0pt S\lower-1pt\hbox{${}
     \scriptstyle'$}}\limits^{\kern2pt\raise.1ex
     \vbox to 0pt{\hbox{$\scriptscriptstyle\rightarrow$}\vss}}}}
\def\vt{\lowfwd t{1.5}1}

\title{Traits and tangles:\\ An analysis of the Big Five paradigm by tangle-based clustering}

\author{Hanno von Bergen and Reinhard Diestel%
   }

\date{\today}

\begin{document}

\maketitle

\begin{abstract}\noindent
Using the recently developed mathematical theory of tangles, we re-assess the mathematical foundations for applications of the five factor model in personality tests by a new, mathematically rigorous, quantitative method. Our findings broadly confirm the validity of current tests, but also show that more detailed information can be extracted from existing data.

We found that the big five traits appear at different levels of scrutiny. Some already emerge at a coarse resolution of our tools at which others cannot yet be discerned, while at a resolution where these {\em can\/} be discerned, and distinguished, some of the former traits are no longer visible but have split into more refined traits or disintegrated altogether.

We also identified traits other than the five targeted in those tests. These include more general traits combining two or more of the big five, as well as more specific traits refining some of them.

All our analysis is structural and quantitative, and thus rigorous in explicitly defined mathematical terms. Since tangles, once computed, can be described concisely in terms of very few explicit statements referring only to the test questions used, our findings are also directly open to interpretation by experts in psychology.

Tangle analysis can be applied similarly to other topics in psychology. Our paper is intended to serve as a first indication of what may be possible.
\end{abstract}

\bigskip\noindent
This paper can be read at two levels.

\begin{itemize}

\item At the methodological level it showcases a new mathematical clustering method suited particularly to the social sciences~\cite{OurTangleSoftware,TangleBook}: indirect clustering by {\em tangles\/}.

\item At the content level it offers additional, non-statistical, soundness validation (which we shall argue is needed) and new structural analysis of the Five Factor Model in personality testing.

\end{itemize}

\smallskip\noindent
   Tangles have their origin in structural graph theory. Although they do offer a new, indirect, approach to generic clustering, which we apply here, their particular strength lies in direct applications to problems typical for the social sciences, as outlined in~\cite{TangleBook}.

\section{Introduction}\label{intro}

The `big five' paradigm, also known as the {\em Five Factor Model\/}~(FFM)~\cite{FiveFactorModel}, holds that the essence of an individual's personality can be captured by evaluating it in terms of the five `OCEAN' traits: {\em o\/}penness to experience, {\em c\/}onscientiousness, {\em e\/}xtraversion, {\em a\/}greeableness and {\em n\/}euroticism. These five traits have been advanced as interpretations of the five principal factors originally found by Tupes and Christal~\cite{Big5Origin} in data of Cattell~\cite{Cattell45} and Fiske~\cite{Fiske49}, and also of five factors%
   \COMMENT{}
   found by others in different data; see~\cite{FiveFactorTheory}.%
   \COMMENT{}
   In a typical personality assessment based on this paradigm, such as~\cite{Neo-PI-R,GoldbergMarkers1992,Big5onlinetest}, the person to be evaluated has to answer, for each of these five traits, a set of questions designed to test to what degree an individual has the trait in question.

There are a number of fundamental assumptions underlying this approach, which are the topic of this paper.

The first of these stems from the fact that the above five traits are not themselves the factors, in the technical factor-analytical sense,%
   \COMMENT{}
   but are interpretations of such factors. In personality tests such as~\cite{Big5onlinetest}, however, they are targeted directly. The multitude of traits studied, e.g., in \cite{Cattell45} and~\cite{Fiske49}, has thus been compressed into five factors; these were interpreted; and their interpretations were fanned out again into the many aspects displayed in~\cite{Big5onlinequestions} and assessed in~\cite{Big5onlinetest}.\looseness=-1

Compression followed by de-compression always bears the risk of information loss or shift. Hence the following question arises:

\begin{itemize}

\item[(1)] To what extent are current personality tests that are designed to target the traits named as OCEAN related, in a reproducible quantitative sense, to the original Five Factors from which they derive their justification?

\end{itemize}

\noindent
We believe that settling this question requires data of a kind we have been unable to find in the literature; see our discussion of this problem in \cref{interpretation1}.%
   \COMMENT{}

Alternatively, one can interpret the reference which these tests make to the original Five Factors as inspiration rather than justification, and seek to validate these tests directly. In particular, do the several questions into which each of the five OCEAN traits is fanned out really test the same thing? For example, we might, in our own use of the term `conscientiousness', count both orderliness and a sense of responsibility for others as aspects of this `trait'.%
   \COMMENT{}
   But do these two phenomena really occur together in individuals more often than not?

Note that this does not follow even if we assume that `conscientiousness' is a suitable name for one of the five principal factors found in those original studies. What this assumption and the definition of factor analysis tell us is that several of the traits originally tested for are highly correlated with {\em some\/} aspect of conscientiousness. It does not tell us that all its aspects we might think of when designing tests such as~\cite{Big5onlinetest} must be highly correlated with each other.\looseness=-1

Ideally, we would like to start with a positive answer to the following question\rlap:\looseness=-1

\begin{itemize}

\item[(2)] Are the five OCEAN traits extensionally verifiable {\em as traits\/},%
   \COMMENT{}
   in some objective sense that does not depend on our own, culturally determined, notions associated with those five words?

\end{itemize}

\noindent
   In other words, can we ground our interpretative notions of the five groups of personality aspects named as OCEAN in quantitatively verifiable data that confirms them as genuine traits, whether or not these correspond exactly to the original Five Factors as stipulated in~(1)?

Note that no experiment can be expected to prove assertions such as `these ten questions are well suited to test for openness' (say). This is because the only way to define a comprehensive trait such as openness rigorously at the data level is by what mathematicians would call a {\em class\/}: to name some explicitly chosen set of `markers' known to be highly correlated with each other, and decree that whatever they have in common (a~postulated entity thought to `explain' their high correlation) shall be called `openness'. But then this set of markers could be tested for directly~-- which would make its justification tautological and offer no added value.

However, one {\em can\/} develop criteria for the following generalisation of~(2), which can also be applied directly to any concrete personality test:

\begin{itemize}

\item[(3)] To what extent does a given set of questions form markers of {\em some\/} trait?

\end{itemize}

\noindent
   Such criteria are suggested in \cref{TraitCriteria}. Their validation involves clustering.

Buchanan, Johnson and Goldberg~\cite{Big5dataPaper} offer evidence of a positive answer to~(3) for a test similar to~\cite{Big5onlinetest}. This evidence is based on factor analysis run on the answers they obtained for questions designed to test for the OCEAN traits. While factor analysis can, in principle, be used for clustering, this is not its brief or even its forte. We shall argue in \cref{TraitsAsClusters} that clustering by factor analysis is too narrowly defined to meet our criteria for~(3) from \cref{TraitCriteria}. If one accepts our criticism, a~consequence will be that the results from~\cite{Big5dataPaper} do not sufficiently support a positive answer to~(3).

Using a new clustering method developed specifically for `fuzzy' datasets as are common in the social sciences, clustering by~{\em tangles\/}~\cite{TangleBook}, we offer here an analysis more in line with our criteria for~(3) of the test results obtained in~\cite{Big5onlinedata1,Big5onlinedata2} for the IPIP questions~\cite{Big5onlinequestions} developed to test the FFM markers of Goldberg~\cite{GoldbergMarkers1992}. We give a preview of this in \cref{setupsection}.

\medbreak

Our analysis broadly confirms that the 50 questions asked in~\cite{Big5onlinequestions} correspond to five traits, in a sense that meets our criteria from \cref{TraitCriteria}. Interestingly, though, these traits are not visible at the same level of `resolution' of our clustering tool of tangles: some already emerge at a coarse resolution at which others cannot yet be distinguished, while at a resolution where these {\em can\/} be distinguished some of the former traits are no longer visible as clusters, but have disintegrated into single questions. {\em Our analysis, thus, finds more structure\/} than just five clusters: structure between the five expected traits, but also between them and their subtraits and more comprehensive traits.

We have found this structure to be robust across different, indeed disjoint, sets of participants whose answers to the questions in~\cite{Big5onlinequestions} we evaluated. But our results are limited, so far, to the data published in~\cite{Big5onlinedata1,Big5onlinedata2}. It would be interesting to see if they can be replicated with results obtained in other personality tests developed for the `big five' paradigm.%
   \COMMENT{}

Tangles have their origin in structural discrete mathematics, in particular in graph theory. The use of tangles in data analysis, including for the social sciences, is new and treated in depth in~\cite{TangleBook}. An easy-to-read introduction can be found in~\cite{TangleBookPreview}, and we outline what we need here in~\cref{tangles}. Tangle software is freely available via~\cite{OurTangleSoftware}.

\subsection{From factors to traits: information loss or shift through interpretation in pursuit of challenge~(1)}\label{interpretation1}

The FFM has its origin in the fact that several studies found five main~factors in personality tests each testing for some larger set of traits, different sets for these original studies. Formally, this precludes any notion that these are `the same' five factors. However, they are said~\cite{FiveFactorTheory} to have similar {\em interpretations\/}: those summarised, e.g., as `OCEAN'. Moreover, only this shift from the data level to that of interpretation enables us to talk about (five) {\em traits\/} at all, rather than about intuitively inaccessible mathematical entities constructed to formally `explain' (linearly combine to) various other traits.%
   \COMMENT{}

When FFM-based personality tests such as~\cite{Big5onlinetest} are developed, however, these interpretations necessarily take on a life of their own: the questions designed to test for the `big five traits' target whatever our intuitive ideas for these traits encompass. In particular, they do not target the entirely abstract five factors that Tupes and Christal~\cite{Big5Origin} found in Cattell's test for 35 other traits.%
   \COMMENT{}

While this is entirely legitimate, and in practical terms unavoidable, it does involve a double jump: one from the original factors to their interpretations, and another from those interpretations to the inventories designed to test them. Any claim that these inventories test for the original Five Factors, therefore, is a hypothesis that needs validation.

In the case of Tupes and Christal's five factors this would require a study in which, for example, the 50-question inventory from Goldberg~\cite{GoldbergMarkers1992} was tested alongside Cattell's original 35~traits. The set of 10~questions targeting a particular one of the five traits, $t$~say, would then have to be shown to be highly correlated to one of the principal five factors found for Cattell's 35~traits in {\em this\/} study, which in turn would have to be shown to be similar in terms of the Cattell questions to the factor found by Tupes and Christal which $t$ interprets.

However, this is not what happened. Instead, Goldberg~\cite{GoldbergMarkers1992} subjected the returns for his 50 questions to another run of factor analysis. As expected, he found five factors: one for each set of 10~questions targeting a trait, and highly correlated to these 10~questions.
This does not amount to a positive answer to our question~(1)~-- only, at best, to a positive answer to our question (2) or~(3).

We look into this further in \cref{TraitsAsClusters}. We shall argue that factor analysis, while being the well-established tool of choice for {\em finding\/} hitherto untargeted fundamental traits in terms of which others can be explained, is not the most powerful tool available for evaluating whether inventories of questions designed to test for these traits live up to their brief. Our aim in this paper is to offer an additional tool for this purpose.

\subsection{$\!\!$What is a trait? Two extensional criteria for (2) and~(3)}\label{TraitCriteria}

Our question~(2) asks whether the five FFM traits are `extensionally verifiable': whether the groupings of personality aspects implicit in our semantic notion of each of the five terms defining these traits are borne out in reality.

For example, there may be reasons for us to subsume certain distinct personality aspects in our notion of agreeableness that do not, in fact, occur more often together in an individual than other aspects. In such a case `agreeableness', while being a convenient term for us to refer to such a {\em combination\/} of personality aspects, would not name a `trait'~-- at least none fundamental%
   \COMMENT{}
   enough to serve as one of five pillars on which to build comprehensive personality assessments.\looseness=-1

The work reported in this paper is based on the assumption that there is a consensus that a {\em trait\/} should satisfy the two formal requirements outlined below.\penalty-200 In \cref{TraitsAsClusters} we describe a method which, we shall argue, can validate these two criteria better than what appears to be the current standard method.%
   \COMMENT{}

Consider a term~$t$ in our language for which we wish to decide whether or not we should call it a `trait'. To decide this, we might devise a questionnaire~$Q_t$ each of whose questions targets one aspect of~$t$ included in our informal notion for~$t$, so that all these aspects are represented in~$Q_t$. 

Our first requirement on $t$ for qualifying as a trait is what we might call {\em extensional cohesion\/}: that the answers received from any sufficiently diverse population for any such%
   \COMMENT{}
   set~$Q_t$ of questions are highly correlated,%
   \footnote{The correlation may be positive or negative. Since positive and negative correlations between two questions can be converted into each other by inverting one of these questions, and positive correlation is transitive if it is high enough, we can make all correlations in a group of pairwise highly correlated questions positive simply by rephrasing some of the questions.}
   \COMMENT{}%
   statistically or in some other relevant sense.%
   \footnote{Tangles in datasets are defined in terms of set partitions rather than pairs of data points. This enables us to apply more comprehensive notions of correlation than statistical correlation of pairs of questions, such as that of `mutual information' from discrete entropy.}
   Only if they are can the answers to the questions in~$Q_t$ given by a concrete person be treated as measuring the same thing~-- the `trait'~$t$. However, such cohesion alone is not enough to warrant~this.

Our second requirement on~$t$ is what we might call {\em extensional completeness\/}. This is defined only when $Q_t$ is one of several sets of questions in some comprehensive questionnaire~$Q$, such as the `big five' inventories in~\cite{GoldbergMarkers1992} and~\cite{Big5onlinequestions}. It stipulates that the answers received for~$Q_t$%
   \COMMENT{}
   shall {\em not\/} be highly correlated%
   \COMMENT{}
   to those received for any set~$Q'\sub Q\sm Q_t$, such as a similar set~$Q_{t'}$ devised to test for a trait~$t'$ other than~$t$. For if they were, and both $Q_t$ and $Q_{t'}$ (say) had already been shown to be extensionally cohesive, then $t$ and~$t'$ would merely be two manifestations of some more comprehensive trait, testable by $Q_t\cup Q_{t'}$ but perhaps not yet understood in interpretation terms. This would raise legitimate concerns about any view of $t$ and~${t'}$ as different personality traits: it might still be convenient for us, but would not be borne out extensionally as envisaged in~(2).\looseness=-1

In this paper we report on our analysis of the data in~\cite{Big5onlinedata1,Big5onlinedata2}, where $Q$ is the set of questions from~\cite{Big5onlinequestions}. This is composed of five subsets~$Q_t$, one for each of the five terms~$t$ in OCEAN. We found that each of these~$Q_t$ is both extensionally cohesive and extensionally complete in~$Q$ at {\em some\/} level of `resolution' of our tools (see \cref{tanglesclustering}). But we also found that this does not happen at the same level: there is no level of resolution for our tangle-based validation tool at which all the five sets~$Q_t$ simultaneously satisfy the above two requirements on traits.\looseness=-1

\subsection{Traits as clusters: a~generic paradigm for~(3)}\label{TraitsAsClusters}

Our two requirements from \cref{TraitCriteria} for sets of variables, in our case questions, to qualify for being regarded as capturing some common `trait' as envisaged in~(3) are best described in clustering terms. 

Let us view the set~$Q$ of all the questions included in a given FFM test as a `space',%
   \footnote{A~metric space, for example~-- we do not need a vector space. Formally, all we need is a set with a positive real-valued function on its pairs.}
   each question $q\in Q$ being a point in this space. If we interpret high correlation between two questions as close proximity in this space, the extensional cohesion and completeness required of each of the five subsets $Q_t$ of~$Q$, one for each trait~$t$, translate to the requirement that these five sets~$Q_t$ {\em should form distinguishable clusters in the space~$Q$.} For example, a~subset of~$Q$ found to be extensionally incohesive could not be a cluster: it might be a union of several clusters not hanging together, or even of proper subsets of such clusters. An extensionally cohesive but incomplete subset of~$Q$ might form only a partial cluster, or there might be no clusters in~$Q$ at all.%
   \COMMENT{}

Buchanan, Johnson and Goldberg~\cite{Big5dataPaper}%
   \COMMENT{}
   and Goldberg~\cite{GoldbergMarkers1992},%
   \COMMENT{}
   offer validations of their FFM test inventories, which are similar to those we investigated \cite{Big5onlinedata1,Big5onlinedata2}. These%
   \COMMENT{}
   can be described in clustering terms, too. In their studies, as well as in \cite{Big5onlinedata1} and~\cite{Big5onlinedata2}, there is an inventory~$Q$ consisting of five sets~$Q_t$ of ten questions, one for each of the five terms~$t$ in OCEAN. In order to validate their test, they ran factor analysis on the returns they received. They found five main factors, one factor~$f_t$ for each of the five sets~$Q_t$, and note that the questions in~$Q_t$ are highly correlated to~$f_t$ but not to any other~$f_{t'}$.%
   \COMMENT{}
   They interpret this as validation of both their test and, presumably, of the FFM itself in the sense of our question~(1).

This should then imply a positive answer also to our weaker question~(3), and indeed to an extent it does. Expressed in clustering terms, their validation shows that the answers received for their tests form clusters around the five factors they found, viewed as virtual centre points: clusters \wrt\ correlation interpreted as proximity, as earlier.

However, this confirmation of~(3) is not the strongest possible which their data might offer. This is because factor analysis, when used for clustering as here, produces clusters subject to constraints that are not needed: it usually%
   \COMMENT{}
   requires the centre points of the clusters to be orthogonal when viewed as vectors in the appropriate inner product space, and the centre points of clusters are chosen so as to maximise proximity%
   \COMMENT{}
   not just to the points in their corresponding cluster, but to {\em all\/} the data points (subject to being orthogonal). This is mitigated somewhat by an application of {\em varimax\/} to the five factors originally found,%
   \COMMENT{}
   but remains an unnecessary constraint from a pure clustering perspective.

	\begin{figure}[htb]
		\center
		\includegraphics[scale=.9]{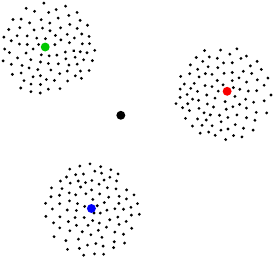}
		\caption{Factor analysis, used for clustering of variables \wrt\ correlation as proximity, might find the black dot as a cluster centre.}
		\label{factorclustering}
	\end{figure}

\cref{factorclustering} shows an extreme example to illustrate the latter point. The small dots indicate the variables to be clustered. Two dots are close in the figure if they are highly correlated as variables. The central black dot indicates a combined variable that factor analysis might find: it is computed so as to maximise its average correlation to {\em all\/} the variables to be clustered, so the black dot is placed so as to minimise its average distance from the small dots.%
   \COMMENT{}
   Standard clustering would instead find the three coloured dots as cluster centres: they each maximise the average correlation only to the group of variables in their respective cluster.\looseness=-1

More generally, recasting our two criteria from \cref{TraitCriteria} in clustering terms indicates that statistical correlation~-- even without the two unnecessary constraints just mentioned~-- is not the only possible basis for their quantification: it is one way to measure the similarity of variables (questions), but not the only one. Even more generally, there may be~-- and indeed are~-- ways of defining and detecting clusters that are not just based on the similarity of {\em pairs\/} of points.

Tangles can detect, and quantify, clusters defined in more general ways than by similarity of pairs of points, let alone just by statistical correlation. Moreover, they unify all standard clustering methods by extracting something like their common structural essence.

\subsection{Testing for our two criteria: from clustering to tangles}\label{TangleClustering}

In \cref{TraitsAsClusters} we argued that generic clustering may offer a way of validating FFM tests that is not subject to some constraints inherent in using factor ana\-lysis for this purpose. In this section we make the case that, among the generic clustering methods available, tangle-based clustering may be a particularly good choice to detect clusters meeting our criteria from \cref{TraitCriteria} in particularly `fuzzy' environments such as personality data.

Clustering, traditionally, seeks to divide a set of data points into subsets of points that are particularly tightly knit together. Sometimes, in {\em soft clustering\/}, points are allowed to spread their `membership' over more than one cluster. But there is no generally accepted definition of `cluster'. This appears to be unavoidable, because the task to decide for every single data point which, if~any, cluster it should be assigned to, is simply too much to ask of real-world data.

Tangles are designed to overcome this problem. They deliberately refrain from attempting to define a cluster as a set of points.  Instead, they merely point to roughly where in the data space clusters can be found, and say something about their relative structure~-- such as whether some large but loose cluster includes one or more smaller dense ones. The gain from ignoring the precise (but error-prone and partly random) information about individual points given in the data, which yields only fuzzy information about cluster-like sets of such points, is that the coarser information retained is precise, robust, and can be examined mathematically. This is what tangles do: {\em tangles are a precise way of capturing intrinsically fuzzy data\/}~\cite{TangleBook}. This applies in particular when the data in question consists of answers to an FFM questionnaire.

The precise but coarser information from the data of~\cite{Big5onlinedata1,Big5onlinedata2} that is encoded in tangles will be enough to answer our question~(3): we do not need to come up with concrete subsets of~$Q$, possibly but not necessarily the sets~$Q_t$, to prove the extensional cohesion and completeness of the five corresponding traits.%
   \COMMENT{}
   And moreover, we obtain additional information on how these traits are related: how they emerge from more comprehensive traits, how they split into subtraits, and all this with quantitative information about the level of precision, or `resolution',%
   \footnote{This will be encoded as the {\em order\/} of the corresponding tangles.}
   at which this happens.

\section{Data, objectives, and setup}\label{setupsection}

Our aim was to validate, and possibly refine, the Five Factor Model (FFM)~\cite{FiveFactorModel} by analysing the returns received in~\cite{Big5onlinedata1,Big5onlinedata2} for the IPIP questions~\cite{Big5onlinequestions} developed to test the FFM markers developed by Goldberg~\cite{GoldbergMarkers1992}.%
   \COMMENT{}
    These questions were designed specifically to test for the five traits commonly abbreviated as OCEAN, ten questions for each of these traits. We applied tangle-based clustering to the set of these questions based only on the answers received, that is, ignoring the information of which trait each question was designed to test.

\subsection{Clusters and traits: the expected outcome}

The anticipated result was that, at the most basic level, we would recover five clusters in this set of questions: the clusters corresponding to, in traditional clustering terms, the five subsets of the questions designed to test for the five traits.%
   \COMMENT{}\looseness=-1

    However, tangle-based clustering can reveal more detailed information than just detecting a certain number of clusters. It can relate these to each other in structural ways which, when we interpret them here, may reveal how those five traits are related: how some of them can be distinguished less clearly than others, and hence form more comprehensive traits at a coarser level, and perhaps split into more detailed traits at a finer level. These levels come on an absolute scale, the {\em order\/} of the tangles found, so that these developments can be compared across the various traits and subtraits as they emerge when the data is analysed at increasing `resolution'. In particular, the five target traits, if recovered, would have their place in any such evolutionary hierarchy of traits that may be factually observable in \cite{Big5onlinedata1} and~\cite{Big5onlinedata2}.

	\begin{figure}[htb]
		\center
		\includegraphics[scale=.95]{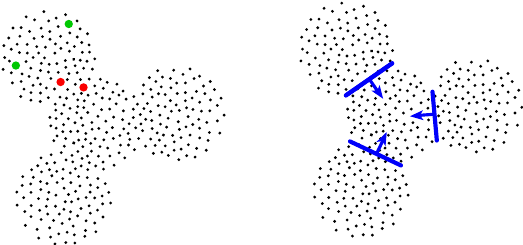}
		\caption{Four clusters and three bottlenecks}
		\label{fourclusters}\vskip-9pt\vskip0pt
	\end{figure}

\subsection{Clustering by tangles}\label{tanglesclustering}

Consider a dataset as depicted in \cref{fourclusters}. Intuitively, there are four clusters: a~central one, and three around it. However as soon as we try to come up with a rigorous explicit definition of `cluster' that underpins this intuition, we get into trouble. For example, if we say that a {\em cluster\/} is a set of pairwise close points, or a maximal such set, we can easily find pairs of points that invalidate this attempt: points that are close to each other but which lie in different point clouds that form our intuitive clusters, such as the two red dots in the picture, or  points that lie in the same intuitive cluster but far from each other, such as the two green dots.

Clustering by tangles sidesteps these issues by taking a radically different, indirect approach. On the right in the picture we can see some natural ways to split the dataset in two, splits that do not cut right through any of the visible clusters but skirt around them. We call such natural partitions the {\em bottlenecks\/} of the dataset. For every bottleneck, every visible cluster lies mostly on one of its two sides: otherwise it would not be a bottleneck. In this way, each cluster {\em orients\/} every bottleneck towards the side that contains most of it. The figure shows this for the central cluster by little arrows on the three bottlenecks.

This information, the simultaneous orientation of each bottleneck towards one of its sides, is called a {\em tangle\/}; it is all that we wish to remember about each of those four clusters. Note that this information is robust in a fuzzy environment such as \cref{fourclusters}: the way in which each of the four clusters orients those three bottlenecks will not change if we draw the bottlenecks slightly differently or move a few of our data points. 

Note, however, that we are still far from a general definition of tangles that can replace the more traditional attempts at defining clusters. This is because our definition of the four tangles in the figure depended on a pre-conceived, if intuitive, traditional notion of cluster: we defined `bottlenecks' as splits of our dataset that leave its clusters largely intact, and we defined `tangles' as simultaneous orientations of all the bottlenecks towards some cluster.

The crucial starting observation in applied tangle theory is that one can define both `bottlenecks' and `tangles' axiomatically, without reference to clusters. One can then either extract point clusters from those tangles~-- or choose not to, if the information about the dataset sought from determining its clusters can be derived from the tangles directly, as will be the case in our application here.

Although tangles contain less detailed information than traditional point clusters, they offer additional insight. One is that they come at different levels: if we allow only few, particularly narrow, bottlenecks we get few tangles that hang together loosely. If we orient wider bottlenecks too, we get more and denser tangles. These will {\em refine\/} the earlier loose ones in that they orient the same bottlenecks as those do, but some wider ones in addition. 

	\begin{figure}[htb]
		\center
		\includegraphics[scale=1]{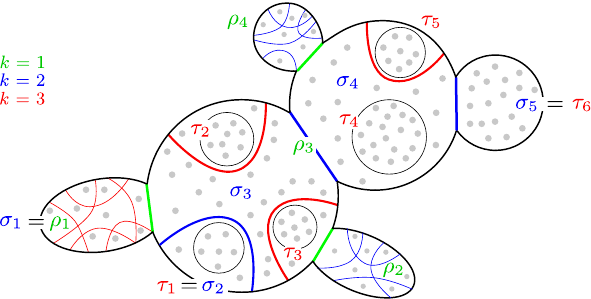}
		\caption{A hierarchy of tangles of increasing order $k=1,2,3$}
		\label{Hierarchy}
	\end{figure}

\cref{Hierarchy}, from~\cite{TangleBook}, shows tangles of increasing {\em order\/}, or bottleneck width. There are four tangles, $\rho_1, \dots, \rho_4$, of order $k=1$. Adding slightly wider bottlenecks to find the tangles of order~2, we see that the tangle~$\rho_1$ of order~1 remains unchanged, the tangles $\rho_2$ and~$\rho_4$ disintegrate, and the tangle~$\rho_3$ splits into the four tangles $\sigma_2,\dots,\sigma_5$. When we allow even wider bottlenecks for tangles of order~3, we find that the tangle $\sigma_1=\rho_1$ disintegrates, the tangles $\sigma_2$ and~$\sigma_5$ remain unchanged, and the tangles $\sigma_3$ and~$\sigma_4$ of order~2 each split into two tangles of order~3, into $\tau_2,\tau_3$ and $\tau_4,\tau_5$ respectively.

The lowest order at which a certain tangle first appears, where it arises as one of several tangles of that order which refine the same tangle of lower order, is known as its {\em complexity\/}; the highest order up to which this tangle remains unchanged is its {\em cohesion\/}, and the difference between the two (plus~1) is its {\em visibility\/}. In the example of \cref{Hierarchy}, the tangle~$\sigma_1$ of order~2 has complexity~1, because it equals (is the unique extension to an order-two tangle of) the tangle~$\rho_1$ of order~1 which refines no tangle of lower order still; it has cohesion~2, because it is still a tangle of order~2 but no longer one of order~3; and so it has visibility $2=(2-1)+1$. The tangle~$\sigma_4$ of order~2 has complexity~2, because it is not the only order-two tangle that refines the tangle~$\rho_3$ of order~1 (as $\sigma_2$, $\sigma_3$, $\sigma_5$ do too); it also has cohesion only~2, since $\tau_4$ and~$\tau_5$ are two tangles of order~3 that both refine it; and so it has visibility $(2-2)+1=1$. See~\cite[Ch.7.4, 14.6]{TangleBook} for details.\looseness=-1

Another advantage over traditional clustering is that the latter often requires us to specify in advance how many clusters we seek to detect. Tangles reflect whatever clusters the dataset contains. We can set a number of parameters, such as%
   \COMMENT{}
   the tangles' required agreement value, which will influence this in general. But once that is done, the number of tangles of any given order found will depend on the data only, not on any choice of ours.

Most importantly, tangles are robust when the data is inherently fuzzy, such as survey data in psychology, sociology, or economics.

\subsection{Tangles and traits}\label{tanglestraits}

When the points in our dataset of \cref{tanglesclustering} are questions about personality, narrow bottlenecks will divide questions whose answer sets are only loosely correlated, in some well-defined sense of our choice.%
   \footnote{As mentioned earlier, the similarity measure between different questions we shall use will mostly be that of mutual information in the sense of information theory, not statistical correlation.}
   As the width of the bottle\-necks considered increases, these bottlenecks (and hence their tangles) can distinguish also between more highly correlated questions. A~tangle will `disintegrate' when the bottlenecks allowed become so wide that they cease to be true `bottlenecks' but dissect the set of questions into singleton subsets (as indicated in \cref{Hierarchy}): when a~question is considered as `highly correlated' only to itself.

In our context it will be possible, in principle,%
   \footnote{Such interpretations need not be readily expressible in traditional terms. Indeed, one of the main advantages of tangles is that they enable us to detect hitherto unknown patterns: patterns we do not have to first guess and then confirm, the standard process in the social sciences. But tangles, once found, are immediately accessible to interpretation by experts; see below.\looseness=-1}
   to interpret each of the tangles of our set~$Q$ of 50 questions found in the data of~\cite{Big5onlinedata1,Big5onlinedata2} as related to one particular personality aspect. These different aspects will be more fundamentally different when their corresponding tangles have low order, because they will be distinguished by splits that cannot separate highly correlated questions. They will be more specific when their corresponding tangles have higher order.\looseness=-1

We can then formalise traits as equivalence classes of tangles of different order, where a~tangle~$\tau$ of order~$k$ is {\em equivalent\/} to a tangle~$\tau'$ of order $k'\ge k$ if, essentially, $\tau'$~is the unique extension of~$\tau$ to a tangle of order~$k'$.%
   \footnote{The precise definition is a little narrower: a tangle $\tau$ of order~$k$ is {\em equivalent\/} to a tangle~$\tau'$ of order~$k'\ge k$ if every tangle of order~$\ell$ with $k\le\ell\le k'$ that induces~$\tau$ is in turn induced by~$\tau'$; see~\cite[Ch.14.6]{TangleBook}.}%
   \COMMENT{}
   A~{\em trait\/}, then, formalised as such an equivalence class of tangles, can be thought of as a single tangle that `is born' at some order~$k$ and `dies' after some maximum order $k'\ge k$.

In \cref{tanglesclustering} we called the minimum and the maximum~$k$ for which `a~given tangle' is a $k$-tangle its `complexity' and `cohesion'. We can now define this more precisely: the {\em complexity\/} and the {\em cohesion\/} of a trait~-- an equivalence class of tangles~-- are the smallest and the largest~$k$ for which this trait (class) contains a tangle of order~$k$. As before, the {\em visibility\/} of a trait is the difference between its cohesion and its complexity plus~1.

Inasmuch as (equivalence classes of) tangles of partitions of a set~$Q$ of questions formalise the informal notion of clusters in~$Q$, our formal notions of complexity and cohesion for traits formalise what in \cref{TraitCriteria} we called the desired external completeness and cohesion of clusters in~$Q$.%
   \COMMENT{}
   They do so in a precise, quantitative, way which can be computed from the data of answer sets received for~$Q$. And they give rise to the new formal notion of visibility, which may be an interesting new parameter for traits found in personality~tests.

\section{Methods 1: Tangle basics}\label{tangles}
For this section let $Q$ be an arbitrary finite set.

\subsection{Partitions}

\begin{definition} {\rm (partition and sides)}\\
Let $A, B \subseteq Q$ be two subsets of $Q$. If $A$ and $B$ are disjoint and $A \cup B = Q$, then $\{A, B\}$ is called a {\em partition of $Q$}. The sets $A$ and $B$ are called the {\em sides} of the partition. 
\end{definition}

We denote partitions either as a set, like~$\{A, B\}$, or by single letters such as $r$ or~$s$. We denote the two sides of a partition~$s$ by $\vs$ and $\sv$. The directions of these arrows are not fixed in advance. For example, if $s=\{A, B\}$ then $A$ could be denoted by either $\sv$ or~$\vs$. But once we have chosen~$A$ to be denoted as~$\vs$, say, the other side~$B$ of this partition will implicitly have been denoted as~$\sv$.

Given a set $S$ of partitions of~$Q$, we denote the set of all their sides as~$\vS$. Note that for every $\vs\in\vS$ also $\sv$ lies in~$\vS$, since it is also a side of~$s\in S$. If~$s$ is a partition of~$Q$ and $q \in Q$, we write $q(s)$ for the side of $s$ that contains~$q$. 

Two sets are commonly called {\em nested} if one contains the other. We call two partitions {\em nested} if they have nested sides.

\subsection{Order functions and similarity functions}\label{OrderFunctions}
\begin{definition} {\rm (order of partition)}\\
 Given a set $S$ of partitions of $Q$ and an $S\to\R$ function $s\mapsto |s|$ that we have called an `order function', we call~$|s|$ the \emph{order} of~$s$, for any $s\in S$. 
\end{definition}

\noindent
   All order functions considered in this paper will take non-negative values. They will also all be injective: not by definition, but because we shall tweak the data so as to make them injective. This makes our computations easier, because it spares us having to make arbitrary choices, but it has no material consequences.

\medbreak

Our aim in choosing order functions will be to assign low order to particularly natural partitions of~$Q$, partitions we called `bottlenecks' in \cref{setupsection}.

One way of defining order functions without reference to any informal notion of cluster is to assign low order to partitions that split few pairs of points we regard as `similar'. Functions $Q^2\to\R$ intended to express this are called \emph{similarity functions} on~$Q$. One simple choice of an order function  on the partitions of~$Q$, then, is to sum up the similarities of the pairs of points split by a partition:

\begin{definition} {\rm (cut weight order)}\\
 Let $w\colon Q^2 \to \mathbb{R}$ be a similarity function on $Q$, and let $\{A, B\}$ be a partition of~$Q$. The {\em cut weight} of $\{A, B\}$ is defined as
$$\text{cut}\,(A, B) := \sum_{a \in A,\, b \in B} w(a, b) \text{.}$$
\end{definition}
\noindent

While the cut weight order function is natural, it tends to assign higher order to more balanced partitions, because these split more pairs. For example, if the sides $A,B$ of a partition of~$Q$ are about equally large, then the number of pairs~$(a,b)$ whose weights $w(a,b)$ are summed in the definition of~{\it cut} is quadratic in~$|Q|$, about $(|Q|/2)^2$, while if the partition is unbalanced in that it divides only some $k$ points of~$Q$ from the rest, this number is linear in~$|Q|$, about~$k\cdot|Q|$. Thus, no matter what the weights~$w(a,b)$ are, the value of ${\it cut}(A,B)$ tends to be larger when $\{A,B\}$ is more balanced.

This is a problem, since we do not want our particularly natural, and hence low-order, partitions to be cluttered up by unbalanced but unnatural partitions. We therefore try to correct the bias of the order function~${\it cut}$ towards unbalanced partitions.  For example, we often take as the order of~$\{A,B\}$ not the sum of the weights~$w(a,b)$, as above, but something more like their average:%
   \COMMENT{}

\begin{definition}\label{DefRatioCutWeight} {\rm (ratio cut weight order)}\\
Given a similarity function on $Q$, the {\em ratio cut weight} of a parti\-tion $\{A, B\}$ of $Q$ is defined as $$\text{Rcut}\,(A, B) := \frac{\text{cut}(A,B)}{|A||B|} |Q| = \text{cut}\,(A,B) \left ( \frac{1}{|A|} + \frac{1}{|B|} \right)  \text{.}$$
\end{definition}

There are many other commonly used order functions; see~\cite{TangleBook}. Our order function of choice will be that of ratio cut weight, based primarily on a similarity function called `mutual information' borrowed from information theory. But we tested also the `cosine' similarity function. Both these are defined in \cref{SimilarityFunctions}.

\subsection{Tangles: basic notion, types, and hierarchy}

Let $S$ be a set of partitions of~$Q$. The few graph-theoretic terms used in this section are explained in~\cite{DiestelBook24}.

\begin{definition} {\rm (orientation of set of partitions)}\\
A set $\sigma  \subseteq \vS$ which, for each $s \in S$, contains exactly one side of~$s$ is called an {\em orientation} of $S$. The side of $s$ that lies in  $\sigma $ is denoted by $\sigma (s)$.
\end{definition}

\cref{fourclusters} indicates an orientation, by arrows, of the set of three `bottle\-neck' partitions shown in blue. As in that example, orientations of partitions can be defined by subsets of~$Q$ that lie mostly on the same side of any such partition. We say that an orientation $\sigma $ of $S$ is {\em guided\/} by any subset $X$ of $Q$ that satisfies
 $$|X \cap \vs| > |X \cap \sv|$$
 for every $\vs \in \sigma $. An orientation $\sigma $ of $S$ can have more than one guiding set, and it can also have none. 

\begin{example} For every $q \in Q$ there is a unique orientation of~$S$ that is guided by $\{q\}$. We call it the {\em principal\/} orientation of~$S$ defined by~$q$.
\end{example}

As explained in \cref{tanglesclustering}, the idea behind tangles is that while clusters, conceived of as subsets~$X$ of~$Q$, are notoriously difficult to pin down precisely, they will~-- whatever their precise definition if any~-- guide the same orientations of the `bottleneck' partitions of~$Q$ if these are narrow enough. For example, if more than two thirds of~$X$ lie on the same side of any $s\in S$, the orientation $\sigma $ of~$S$ guided by~$X$ will not change if we modify~$X$ a little, perhaps by using a competing definition of point cluster, or because our data is not perfectly reliable.\looseness=-1

In this example, the intersection of any three sets in~$\sigma $ contains an element of~$X$, since each of them misses less than a third of~$X$. This motivates the following formal definition of tangles:

\begin{definition} {\rm (tangle)}\\
An orientation~$\tau$ of~$S$ is a {\em tangle\/} of~$S$ if $\vr\cap\vs\cap\vt\ne\es$ for all triples $\vr,\vs,\vt$ of elements of~$\tau$.
\end{definition}

Note that this definition, while motivated by the properties of a subset~$X$ of~$Q$ that we might think of as a point cluster, no longer depends on such an informal notion of `cluster': the point it requires to lie in the intersection of any three%
   \footnote{There is some magic in this number, which is discussed at length in~\cite[Section\,7.3]{TangleBook}.}
   elements of a tangle can be any point in~$Q$ and does not have to come from any previously defined `cluster'~$X$.

Due to this difference,  a set of partitions of a dataset~$Q$ can have tangles that are not guided by any obvious subsets of~$Q$: tangles are more general than orientations of set partitions guided by `clusters' in that set, however defined.

In order to reduce the number of tangles of~$S$, we often require that its triples contain not just one but some specified number of elements of~$Q$:

\begin{definition} {\rm (agreement value)}\\
The {\em agreement value} $a(\tau)$ of a tangle~$\tau$ of~$S$ is the minimum number of elements of~$Q$ that lie in the intersection of any three sets in~$\tau$.
\end{definition}

Thus, if every three sides of partitions of~$S$ that lie in a given tangle~$\tau$ have at least $n$ elements of~$Q$ in common, the agreement value of~$\tau$ is at least~$n$.

Recall that every $q\in Q$ defines a map $s\mapsto q(s)$ from $S$ to~$\vS$, just as every tangle~$\tau$ of~$S$ defines an $S\to\vS$ map $s\mapsto\tau(s)$. We think of~$q$ as being similar to~$\tau$ if these two functions agree on many $s\in S$:

\begin{definition} {\rm (point-tangle similarity)}\label{point-tangle_sim_def}\\
Given a tangle~$\tau$ of~$S$ and any $q\in Q$, we call the number
 $$\sigma(q,\tau) := \big|\{\,s\in S : q(s) = \tau(s)\,\}\big|$$
 the {\em similarity\/} between $q$ and~$\tau$.
\end{definition}

\begin{definition}\label{DefSk} {\rm ($S_k$; $k$-tangles of~$S$; tangles in~$\vS$; order of a tangle; $\T(S)$)}\\
Let $s\mapsto |s|\in\R$ be an order function on~$S$. For every integer~$k$ we write
 $$S_k := \{s \in S: |s| < k\}$$
 and call the tangles of~$S_k$ the {\em $k$-tangles of}~$S$ or the {\em tangles of order~$k$ in}~$\vS$. We write~$\T(S)$ for the set of all tangles in~$\vS$, irrespective of their order. 
\end{definition}

Note that the $k$-tangles of~$S$ are not tangles of~$S$ (unless $k$ is very large), since they only orient its subset~$S_k$. For $k < k'$, every $k'$-tangle~$\tau'$ of~$S$ defines (or {\em induces\/}, or {\em refines\/}) a $k$-tangle of~$S$, the tangle $\tau := \tau'\cap\vSk$ of~$S_k$. Note that $\tau\sub\tau'$ in this case. Conversely, given any two tangles $\tau\sub\tau'$ in~$\vS$, there will be integers $k\le k'$ such that $\tau'$ is a $k'$-tangle of~$S$ and $\tau = \tau'\cap\vSk$ a $k$-tangle.%
   \COMMENT{}

\begin{definition} {\rm (tangle equivalence)}\label{tangleequivalence}\\
 A $k$-tangle $\tau$ of~$S$ is called {\em equivalent\/} to a $k'$-tangle~$\tau'$ of~$S$ with $k\le k'$ if every $\ell$-tangle of~$S$ with $k\le\ell\le k'$ that induces~$\tau$ is in turn induced by~$\tau'$.
\end{definition}

   It is not hard to prove, but also not entirely obvious, that this is indeed an equivalence relation on the set of tangles in~$\vS$. Note that refinement is well defined on equivalence classes: if a tangle~$\tau$ refines a tangle~$\sigma$ not equivalent to~$\tau$,%
   \COMMENT{}
   then any tangle~$\tau'$ equivalent to~$\tau$ refines any tangle~$\sigma'$ equivalent to~$\sigma$.

\begin{definition} {\rm (complexity, cohesion, visibility)}\label{visibility}\\
 The {\em complexity\/} and the {\em cohesion\/} of an equivalence class of tangles in~$\vS$ are the smallest and the largest order of a tangle in that class, respectively. The {\em visibility\/} of the class is its cohesion minus its complexity plus~1.%
   \COMMENT{}
\end{definition}

\begin{definition} {\rm (tangle search tree)}\label{TangleSearchTree}\\
 The {\em tangle search tree} of $S$ is the graph with node set~$\T(S)$ in which two nodes $\rho,\tau$ are adjacent if $\rho \subsetneq \tau$ and there is no tangle $\sigma$ in~$\vS$ such that $\rho \subsetneq \sigma \subsetneq \tau$. We often write~$\T(S)$ also for this tree, not just for its set of nodes.
\end{definition}

The tangle search tree is indeed a tree. Since $S_k = \es$ for sufficiently small~$k$, and $\es$ is a $k$-tangle for this~$k$, the empty set~$\es$ is a node of~$\T(S)$, which we take as its root. If our order function is injective, as we shall make it throughout, then every child of a tangle in the tangle search tree orients exactly one more partition in~$S$ than its parent, so every node has at most two children. The tangle search tree has a non-root level for every integer~$k$ that occurs as an order of a partition in~$S$. If these are all the integers between~1 and some~$K$, the tangles at level $k = 1,\dots,K$ are the $(k+1)$-tangles of~$S$, those that orient every partition in~$S$ of order up to~$k$.

Note that two tangles in~$\T(S)$ are equivalent if and only if the path that links them in the tangle search tree is vertical%
   \footnote{Formally: ascending or descending in the tree order on~$\T(S)$ associated with the root~$\es$.}
   and has no inner vertices at which the tree branches: in graph-theoretical terms, its inner vertices all have degree~2. Contracting the maximal such paths in the tree~$\T(S)$ to single edges between their ends, and deleting the root~$\es$ if it has only one child,%
   \COMMENT{}
   turns~$\T(S)$ into a {\em binary\/} tree~$\hat\T(S)$, one whose non-leaf nodes%
   \footnote{The root of a tree never counts as a leaf~\cite{DiestelBook24}.}%
   \COMMENT{}
   all have exactly two children. These nodes correspond to the equivalence classes of tangles defined earlier.%
   \COMMENT{}

\begin{definition} {\rm (tree of traits)}\label{TreeOfTraits}\\
In contexts when the equivalence classes of tangles in~$\vS$ $($other than possibly~$\{\es\})$%
   \COMMENT{}
   are called `traits', the tree on~$\hat\T(S)$ will be called the {\em tree of traits}. We denote this tree by~$\hat\T(S)$~too.
\end{definition}

For every pair of tangles in~$\vS$ of which neither induces (i.e., is a superset of) the other there exists an $s\in S$ which these two tangles orient differently. We say that $s$ {\em distinguishes\/} these two tangles. If $s$ has lowest order among all the partitions in~$S$ that distinguish the two tangles, it distinguishes them {\em efficiently\/}.

\begin{definition} {\rm (efficient distinguisher of incomparable tangles)}\label{efftdist}\\
If the order function on~$S$ is injective, then any $s\in S$ that distinguishes two tangles in~$\vS$ efficiently is unique. We call it their \emph{efficient distinguisher}.
\end{definition}

\section{Methods 2: Tangle particulars for our analysis}%
    \label{sec:methods2}

Our aim in this paper is to determine, and analyse, tangles of partitions of the set~$Q$ of questions used in the FFM test in~\cite{Big5onlinedata1,Big5onlinedata2}. We shall always denote as~$I$ the set of individuals on whose answers to these questions we base our tangles. This set~$I$ will vary as we compute and test our tangles, as will the set~$S$ of partitions of~$Q$ whose tangles we compute. But $Q$ will always be the same. Thus, we shall study the tangles in~$\vS$ for a suitable set $S$ of partitions of~$Q$, based on the answers to~$Q$ given by various sets~$I$ of people.

Before we describe more specifically how we chose the parameters giving rise to these tangles, let us formalise the notion of `trait' as envisaged in \cref{tanglestraits}, now rigorously on the basis of \cref{tangleequivalence}:

\begin{definition} {\rm (tangle trait)}\label{trait}\\
A  {\em (tangle) trait\/} is an equivalence class of tangles in~$\vS\!$ other than~$\{\es\}$.
\end{definition}

\noindent
   Note that $\{\es\}$ may or may not form an equivalence class in~$\T(S)$: it does so if and only if the empty tangle~$\es$ has two descendents in the tangle search tree.

\medbreak

In order for the term of `tangles in~$\vS$' to be defined, we had to choose an order function on~$S$. We used the order function of `ratio cut weight' introduced in \cref{OrderFunctions}. This order function requires us to first choose a similarity function on~$Q$. We used two different options for this, in order to test our results for stability. These similarity functions are defined in \cref{SimilarityFunctions}.

Whichever similarity function on~$Q$ we worked with, a~key element of our work was that these similarities should be extensionally determined: they should be computable from how the individuals in~$I$ answered those questions, regardless of any interpretation, or intention for why those questions were included in~$Q$, such as to test for a particular trait. Moreover, our aim was that every individual in~$I$\penalty-200\ should carry the same weight in determining the similarities between questions in~$Q$. Since different individuals have different habits in how they answer {\em any\/} question, such as their different levels of assertiveness or positivity, our aim of giving them equal weight required us to factor out those individual habits. We describe how we did that in \cref{NormalisingAnswers}. Note that such normalisation must be done with care, since the personal habits we normalise are themselves personality traits like those that $Q$ intends to measure. We also normalised the answers for each particular question to give them mean zero, so that questions suggesting traits generally seen as positive were treated no differently from other questions.

In \cref{ChoosingPartitions} we explain how we chose~$S$. In \cref{FindingTheTangles} we present our tangle search algorithm, slightly simplified; the precise version is described in~\cite{OurTangleSoftware}.

\subsection{\boldmath Normalising the answers received for~$Q$}\label{NormalisingAnswers}

In our data, the answers solicited by the questions in~$Q$ were on scale from 1 to~5, with 1 indicating strong disagreement and 5 strong agreement. This makes $Q$ into a set of functions $q\colon I\to\R$, where ${q(i)\in\{1,\dots,5\}}$ records the answer given to~$q$ by~$i$. Dually, every $i\in I$ defines a function $i\colon Q\to\R$ that sends each question~$q\in Q$ to its answer~$i(q)$ given by~$i$. Thus, $i(q)=q(i)$ for all $i\in I$ and $q\in Q$ whenever some~$I$ has been fixed.

Some individuals tend to agree more than others with questions as they are phrased. This can be seen formally by the fact that the functions $i\colon Q\to\R$ have different means~$\mu_i$ for different $i\in I$. To compensate for this we normalised, for every $i\in I$ separately, their answers to give them mean~0, by subtracting~$\mu_i$ from their given answers.

Next, we compensated for different levels of assertiveness amongst different people. The assertiveness of person $i\in I$ can be measured by the standard deviation~$\sigma_i$ of the function $i\colon Q\to\R$. We normalised these by dividing, for each $i\in I$ separately, their answers (after normalising their mean to~0) by~$\sigma_i$, to give them all the standard deviation of~1.

In a third step, we did some normalisation between the questions ${q\colon\! I\!\to\R}$. Some questions have more people agree with them than other questions, perhaps because they are testing for a trait that is generally seen as positive. We compensated for this by subtracting, separately for each question $q \in Q$, from the answers received for~$q$ the median of all these answers.

After this last normalisation, every $q\in Q$ had equally many answers $i(q)\le 0$ as answers $i(q)\ge 0$, counted over all $i\in I$. However, this last normalisation between the questions may have undone some of our earlier normalisation between individuals. We therefore repeated all three steps cyclically until this process converged to functions that satisfied all three of our normalisation goals: to functions $i\colon Q\to\R$ and $q\colon I\to\R$ with $i(q)=q(i)$ such that every~$i$ has mean~0 and standard deviation~1, and every~$q$ has median~0. On the data we investigated, this convergence happened quickly, and only a few rounds of iteration were needed.

From now on we shall use our notation of $q\colon I\to\R$ for~$q\in Q$, and $i\colon {Q\to\R}$ for~$i\in I$ to refer to these normalised functions computed from their original versions in the data of~\cite{Big5onlinedata1,Big5onlinedata2}.

\subsection{Similarity functions}\label{SimilarityFunctions}

Our first similarity function is based on the notion of discrete entropy in information theory, defined below. More background, examples, and a discussion of how this captures similarites can be found in \cite[Ch.\,9.5]{TangleBook}.

\begin{definition} {\rm (entropy and mutual information)}\\
The \emph{(discrete) entropy} of a function $f\colon I \to X$, where $X$ is any finite set, is~the number
 $$\H(f) := \sum_{x \in X} \big({|f^{-1}(x)|}/{|I|}\big)\cdot \log\! \big({{|I|}/|f^{-1}(x)|}\big) \text{.}$$
The \emph{product} $(f \times g)\colon I \to X \times Y$ of two functions $f: I \to X$ and $g: I \to Y$ is defined by setting $(f \times g)(i) := (f(i), g(i))$ for all $i\in I$.
The \emph{mutual information} of $f: I \to X$ and $g: I \to Y$, then, is defined as
 $$ \I(f, g) := \H(f) + \H(g) - \H(f \times g) \text{.}$$%
   \COMMENT{}
\end{definition}

\begin{definition} {\rm (entropy similarity)}\label{entsimdef}\\
The \emph{entropy similarity} of two questions $p,q\in Q$ is their mutual information $\I(p,q)$ as $I\to\R$ functions.
   \end{definition}%
   \COMMENT{}

If we view the set $\R^I$ of all $I\to\R$ functions as a copy of the vector space~$\R^n$ with $n = |I|$, we can compare entropy similarity to other similarity functions between pairs of points of~$\R^n$. We shall do this with the following similarity measure on~$\R^n$, which is based on the cosine of the angle between two vectors:

\begin{definition}\label{cosimdef} {\rm (cosine similarity)}\\
The {\em cosine similarity} of two points $p, q \in \R^n$ is defined as $$c(p, q) := \frac{p \cdot q}{{\Vert}p{\Vert} \text{ } {\Vert}q{\Vert}} \text{.}$$%
   \COMMENT{}
Here $\cdot$ denotes the dot product in~$\R^n$, while ${\Vert}p{\Vert}$ and~${\Vert}q{\Vert}$ denote the lengths of $p$ and~$q$, respectively. 
The {\em absolute cosine similarity} of $p$ and $q$ is defined as
  $$ac(p, q) := |c(p, q)| \text{.}$$
\end{definition}

\subsection{Choosing partitions}\label{ChoosingPartitions}
As $|Q|=50$, there are about $2^{50}$ partitions of~$Q$,%
   \COMMENT{}
   so we could not include them all in our set~$S$ of partitions whose tangles we computed. This set $S$ should contain a diverse selection of low-order partitions, so that our results approximated the tangles of the set of all partitions of~$Q$.  

We constructed such a set~$S$ in three steps, described below. Performing these steps entailed two choices: of one of two matrices, $L$ or~$J$ (see below); and of one of the two similarity measures from \cref{SimilarityFunctions}.%
   \COMMENT{}
   As our order function $s\mapsto |s|$ we always used `ratio cut weight' from \cref{DefRatioCutWeight}.

In the first step we computed an orthonormal set of eigenvectors of a certain matrix, which we then converted into partitions of~$Q$. This yielded a set of at most $|Q|=50$ partitions, typically fewer. We did this for one of two $m\times m$ matrices, $L$~and~$J$, where $m := |Q| = 50$. Let the rows and columns of these matrices be indexed not by numbers but by the elements of~$Q$ directly.

The matrix~$L$ is commonly known as the `combinatorial Laplacian' of the complete graph on~$Q$ with `edge weights' given by a similarity function~$\sigma$ on~$Q$. In row~$p$ and column~$q$ the matrix~$L$ has the entry~$-\sigma(p,q)$ if $p\ne q$, and entry $\sum_{q'\in Q\sm\{q\}}\sigma(q,q')$ if $p=q$. See \cite[Ch.\,9.2,\,10.3]{TangleBook} for more about~$L$ and the role of its eigenvectors.%
   \COMMENT{}
   The similarity functions we used are those from \cref{SimilarityFunctions}.

$L$~has $m$ real eigenvalues $0=\lambda_0\le\dots\le\lambda_{m-1}$, counted with multiplicities. If~$Q$ had a partition~$\{A,B\}$ such that $\sigma(a,b) = 0$ for all $a\in A$ and $b\in B$, we would add it to~$S$ straight away and start again with $Q$ replaced by~$A$ and, separately, by~$B$. This did not in fact happen. The first eigenvalue $\lambda_0 = 0$ of~$L$ therefore always had multi\-plicity~1 (as is well known),%
   \COMMENT{}
   so $0 < \lambda_1\le\dots\le\lambda_{m-1}$.%
   \COMMENT{}
   We then picked eigenvectors $v_1, v_2,\dots$ for the eigenvalues $\lambda_1,\lambda_2,\dots$ one by one, all orthogonal to both the eigenvector $\mathbbm 1$ for~$\lambda_0$ and to each other.

The matrix~$J$ has entries $-(p\cdot q)$ in row~$p$ and column~$q$, for all $p,q\in Q$, where `$\cdot$' once more denotes the dot product in~$\R^n$, as in \cref{cosimdef}.%
   \COMMENT{}
   The matrix~$J$ has $m$%
   \COMMENT{}
   real eigenvalues $\lambda_1\le\dots\le\lambda_m\le 0$, counted with multiplicities, whose corresponding orthogonal eigenvectors $v_1, v_2,\dots$ are the {\em principal components\/} of the `variables' in~$Q$ for the data given by~$I\sub\R^Q$.%
   \COMMENT{}
   Note, however, that we are using these principal components not as discussed, and criticised, in \cref{TraitsAsClusters} (see \cref{factorclustering}), but in the dual (and opposite) role: rather than proposing them as centre points of clusters in~$Q$, we use them to define natural partitions of~$Q$.

The eigenvectors $v_1, v_2,\dots$ of either $L$ or~$J$ were then turned into partitions of~$Q$. Given $v_i\colon Q\to\R$, we chose the partition
 $$s_i := \big \{ \{q \in Q: v_i(q) < 0\}, \{q \in Q: v_i(q) \geq 0\} \big \} \text{.}$$
   Let us call the set~$S$ of these partitions~$S^0$.

In the second step we iteratively enlarged our current set~$S$ of partitions by including, for all the pairs of partitions $\{A, B\}, \{C, D\}$ already in~$S$, also their four {\em corners} in~$\vS$. These are the sets
 $$A \cap C,\ B \cap C,\ A \cap D,\ B \cap D.$$
 The partitions of~$Q$ which these sets form with their complements in~$Q$ are the four {\em corner partitions\/} of $\{A, B\}$ and~$\{C, D\}$.
Tangle algorithms, see \cite{OurTangleSoftware}, work better when corners of partitions in $S$ are also in $\vS$ if they have low order. Let us call the resulting set~$S$ of partitions~$S^1\supe S^0$.%
   \COMMENT{}

We then performed the third step on the partitions in~$S^1$. For each $s\in S^1$ we tried to find a single~$q\in Q$ such that moving $q$ across the partition~$s$ would reduce the order of~$s$. If we were able to find such elements~$q$, we moved the~$q$ across~$s$ for which this decreased the order of~$s$ the most, adding the modified partition to~$S$.%
   \COMMENT{}
   We then repeated this with the modified partition, iteratively until no such~$q$ could be found.%
   \COMMENT{}
   We added all these modified partitions to~$S^1$.%
   \COMMENT{}
   We took the resulting set of partitions of~$Q$ as the set~$S$ whose tangles we would finally compute.

Let us take a moment to motivate the above three steps; more background can be found in~\cite[Section~11]{TangleBook}. Partitions obtained from orthonormal eigenvectors of $L$ and~$J$ are particularly useful starting partitions for our construction: they are more balanced than other partitions of low order, in that they divide $Q$ into sides of roughly equal size, and among such balanced partitions they have particularly low order.

Adding corners to a given set~$S$ of partitions, such as~$S^0$,%
   \COMMENT{}
   is easily seen not to place additional constraints on its tangles when we require those to extend to tangles of the enlarged set. But it helps with weeding out `fake' tangles of the given set~$S$: $k$-tangles that do not extend to $k$-tangles of all the partitions of~$Q$, those we are trying to approximate for each~$k$.

Adding partitions obtained by moving single elements across an existing partition has a similar effect.%
   \COMMENT{}
   It can weed out $k$-tangles of~$S^1$ that fail to extend to a $k$-tangle of all the partitions of~$Q$, but requiring tangles of~$S^1$ to extend to tangles of the enlarged set~$S$ places no additional demands on its good tangles. This is because tangles of agreement~2 cannot orient two partitions differently if these differ by a single $q\in Q$, i.e., both towards the moving~$q$ or both away from it.
    Hence the way in which a tangle can orient the additional partitions in~$S$ is determined by how it orients~$S^1$: each tangle of~$S^1$ extends to at most one tangle of~$S$. And those that do not extend are rightly weeded out, since they satisfied the tangle axioms `for the wrong reason': because $S^1$ did not yet include enough low-order partitions to put the tangle axioms to a sufficiently rigorous test. See \cite[Section~11.1--2]{TangleBook} for details.

\subsection{Finding the tangles}\label{FindingTheTangles}

Having constructed one of the sets~$S$ of partitions of~$Q$ described in \cref{ChoosingPartitions}, we constructed all the tangles in~$\vS$ of agreement value at least~$a=2$, as follows.

We began by sorting the partitions in~$S$ by order, as $|s_0| < |s_1| < \ldots $ say. Then for $k=0,1,\dots$ in turn we built the set~$\T_k$ of all tangles of~$\vSk$, where
 $$S_k := \{s_i\mid i < k\}.$$
 Note that this coincides with our \cref{DefSk} of~$S_k$ if $|s_i|=i$ for all~$i<k$, which we shall assume to simplify our exposition.%
   \footnote{The actual value of an order never matters for tangles: all that matters is the linear ordering which the order function used imposes on~$S$.}
   Note that $\T_0 = \{\es\}$, since the empty tangle is the unique tangle of $S_0=\es$.

Having constructed~$\T_k$ for some~$k\ge 0$, we let~$\T_{k+1}$ consist of all the tangles of the form $\tau\cup\{\vsk\}$ or $\tau\cup\{\svk\}$ with $\tau\in\T_k$. For any given~$\tau\in\T_k$, sometimes both these extensions of~$\tau$ were tangles, or just one of them, or neither: this depended on whether $\vsk$ or~$\svk$ formed a triple with two elements of~$\tau$ whose intersection consisted of fewer than~$a$ elements of~$Q$, in which case adding it to~$\tau$ would not yield a tangle of~$S_{k+1}$ of agreement value at least~$a$.

This process ended only once the last partition in~$S$, its element of largest order, was examined. Then $\T := \T_1\cup\T_2\cup\dots$ was our set of all tangles in~$\vS$.

\section{Results}\label{results}

In this section we present our findings. In \cref{traitsfound} we specify the datasets in which we looked for tangles. We state which parameters we used to compute these tangles (see \cref{SimilarityFunctions,ChoosingPartitions}), and note how many traits they yielded (\cref{trait}). We finally define what we mean when we use the `big five' labels of $O,C,E,A,N$ to name some of these traits. 

In \cref{traitstructure} we present the trees of traits we found (\cref{TangleSearchTree,TreeOfTraits}), analyse their structure, and indicate what this means from an interpretation point of view.

Interpretation of our results is addressed more comprehensively in \cref{interpretation}. While we have to leave interpretation as such to experts in psychology, we say how exactly such experts can read off the facts they need from our structural analysis in \cref{traitstructure} and supplementary data listed in the Appendix. One of the fortes of tangle clustering is that it condenses the most crucial information into a very small set of facts that can be stated explicitly at the interpretation level. This is done in the Appendix, in \cref{distinguishers}.

In \cref{point-tangle-sim} we list, for each of the traits found, which of the 50 questions in~$Q$ represent that trait best. In particular, we do this for the $OCEAN$ traits, but also for the other traits we found. This also has a direct impact on interpretation, which is discussed in \cref{point-tangle-sim} too. 

Finally, we address the question of how reliable our results are. This has two aspects, which we call `robustness of methods' and `stability of findings'. 
By {\em robustness\/} of our methods we mean invariance of our findings under small changes of the input, while we keep those methods unchanged. By {\em stability\/} we mean the converse: the invariance of our findings under small changes of the methods, performed on the same input. These are discussed in \cref{robustness,stability}, respectively.

\subsection{The tangle traits we found}\label{traitsfound}

We computed, separately for the two datasets in \cite{Big5onlinedata1} and~\cite{Big5onlinedata2}, the tangles in~$\vS$ of agreement at least~2. We focus here on reporting our findings for the larger of those two datasets, the data from~\cite{Big5onlinedata2}. We also report more briefly what tangles we found in~\cite{Big5onlinedata1}, and discuss the differences (\cref{traitstructure}) and their significance.%
   \COMMENT{}
   Both studies were conducted with the same questionnaire~$Q$ of 50 questions. When computing their tangles we used identical parameters, specified below.

Let us remark right away that there will indeed be differences between our findings for the two studies of \cite{Big5onlinedata1} and~\cite{Big5onlinedata2}. This motivated us to investigate to what extent these differences were due to substantial differences in the data, or perhaps to a lack of robustness of our~methods. To answer this we re-applied our algorithm, with the same parameters as before, to disjoint subsets of the participants in the larger study of~\cite{Big5onlinedata2}. As we shall see in \cref{robustness}, the tangles we found on these subsets were very similar to those we found on the entire dataset of~\cite{Big5onlinedata2}. Hence they all differed, and in similar ways, from the tangles we found in~\cite{Big5onlinedata1}.

We have thus been able not only to find reproducible%
   \COMMENT{}
   tangle-based traits in the studies of \cite{Big5onlinedata1} and~\cite{Big5onlinedata2}. We can also say with some confidence which aspects of the relationship between those traits, in each of the two datasets, are stable across those two studies (which were four years apart and conducted with different target sets of participants), and which were specific to one study or the other.

The tangles analysed below are based on mutual information as the similarity function (\cref{entsimdef}), and on its Laplacian~$L$ as the matrix (\cref{ChoosingPartitions}).%
   \COMMENT{}
   In \cref{stability} we shall compare these tangles with the tangles we found for the cosine similarity function (\cref{cosimdef}), again with the matrix~$L$. Since these two similarity functions measure different things, we expected them to return different tangles, which was indeed the case. As both mutual information and cosine similarity are accessible to interpretation, the (small but interesting) difference between the tangles found can also be interpreted. We must leave this to the experts in psychology, but would like to point out this possibility.

We further ran tests to see whether our choice of the matrix~$L$ rather than~$J$ affected the tangles we found. Any difference here would be harder to interpret. However, we found no such difference; see \cref{stability}.

In each of the datasets from \cite{Big5onlinedata1} and~\cite{Big5onlinedata2} we found several thousand tangles.%
   \COMMENT{}
   In the case of~\cite{Big5onlinedata1} these combined to eleven traits, in the case of~\cite{Big5onlinedata2} to fifteen (see \cref{{trait}}). We denoted some of these traits by the letters $O, C, E, A, N$, according to the following rule.

Let us first denote by the letters $O, C, E, A, N$ the five subsets of~$Q$ designed to test for the trait indicated by the respective letter. Consider any one of these subsets, $O$~say. If it so happened that $O$, but none of the other four sets~$C,E,A,N\sub Q$, guides the lowest-order tangle in some trait~$T$, and if no trait refined by~$T$ has this property too, we denoted~$T$ as~$O$.%
   \COMMENT{}
   Similarly, we used each of the letters $C,E,A,N$ to denote the set of questions in~$Q$ it indicates, and then by implication also the trait corresponding to that~set, as defined above for~$O$.

The tree of the fifteen traits we found in~\cite{Big5onlinedata2} is depicted in~\cref{maintst}. The tree of the eleven traits we found in~\cite{Big5onlinedata1} is depicted in~\cref{smalltst}.

\subsection{Structure: how different traits are related}\label{traitstructure}

\begin{figure}[htb]
	\center
	\includegraphics[scale=0.63]{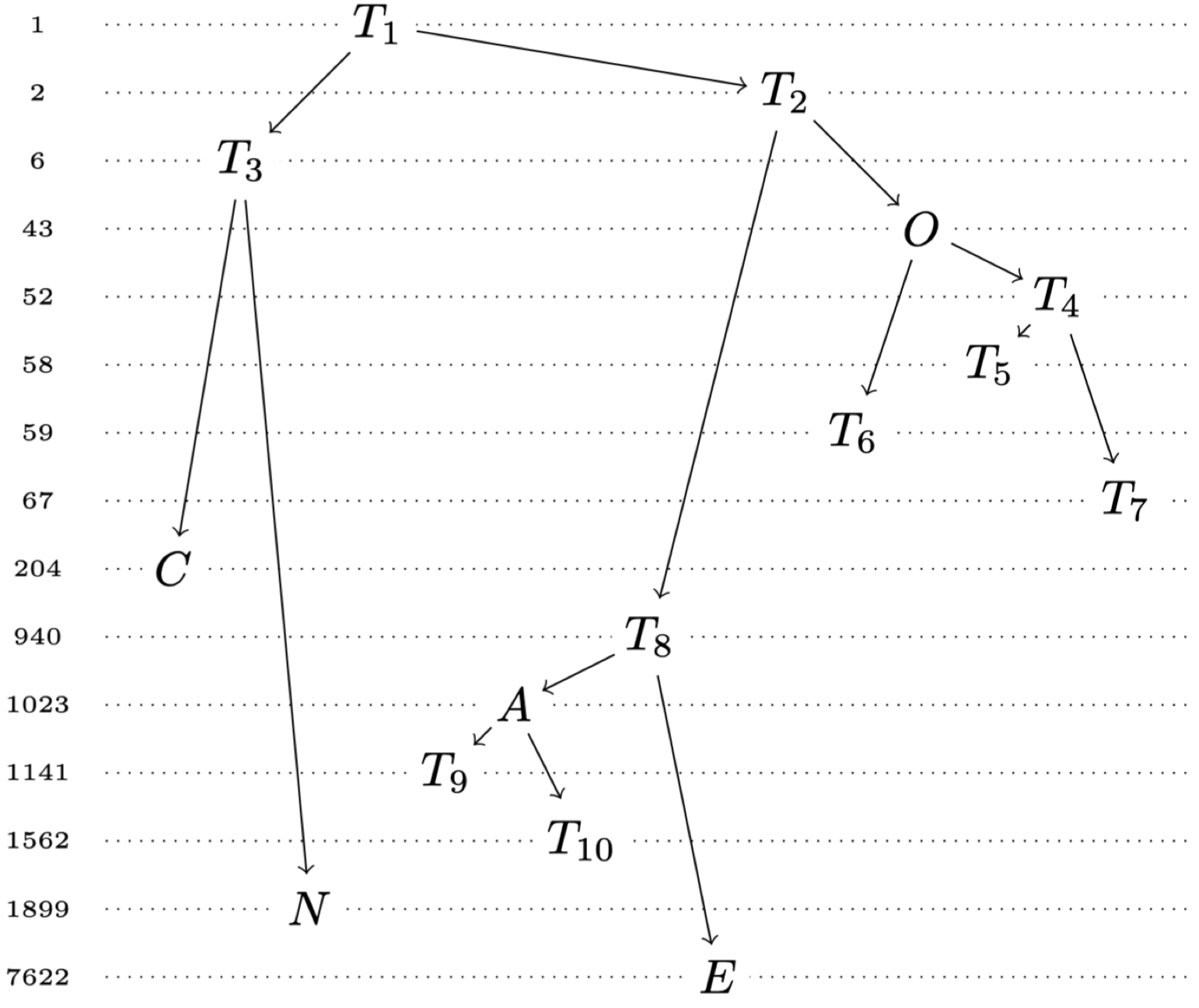}
	\caption{The tree of traits found in the larger study of~\cite{Big5onlinedata2}}\vskip-9pt\vskip0pt
	\label{maintst}
\end{figure}

\begin{figure}[hbt]
	\center
	\includegraphics[scale=0.55]{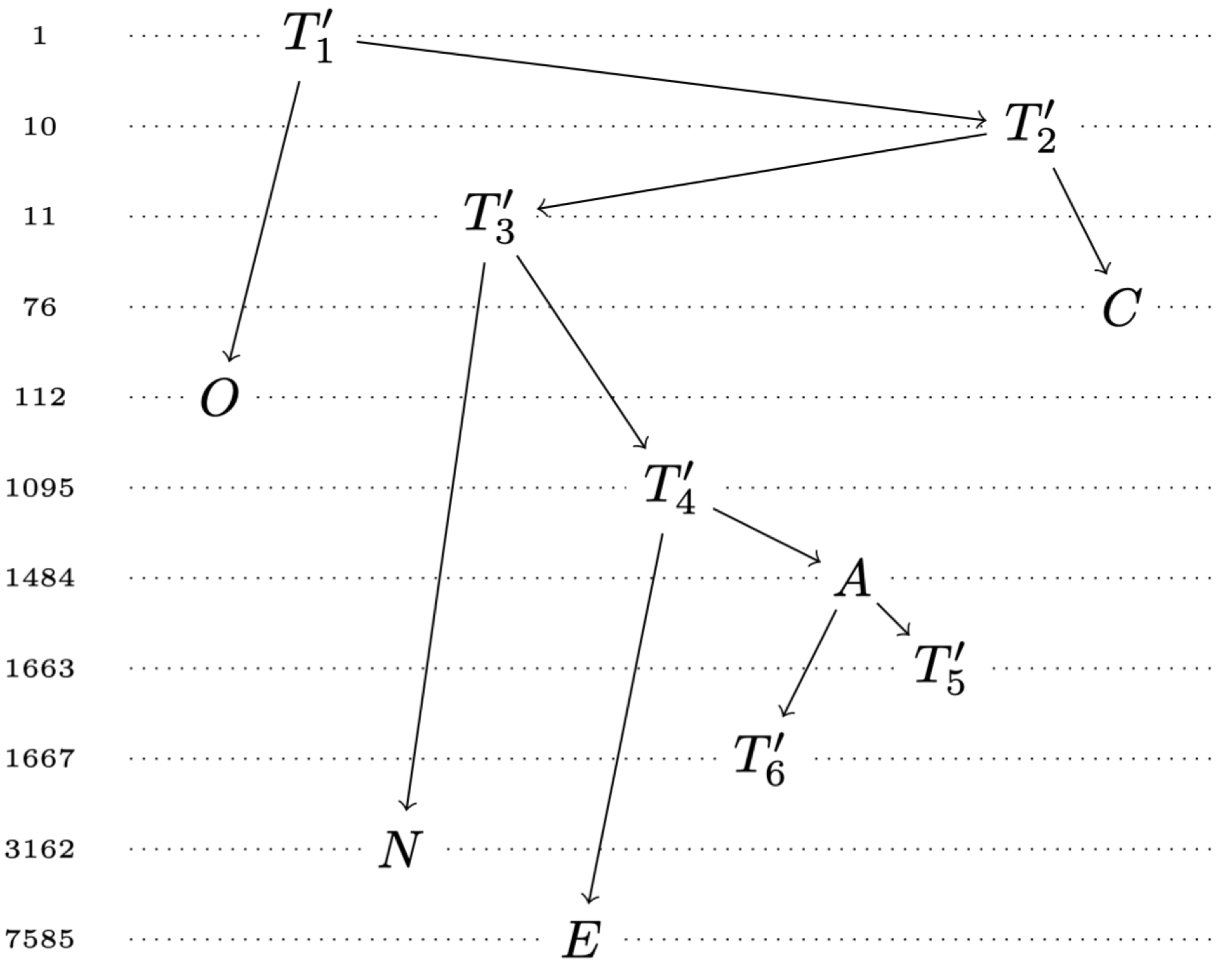}
	\caption{The tree of traits found in the smaller study of~\cite{Big5onlinedata1}}\vskip-3pt\vskip0pt
	\label{smalltst}
\end{figure}

The trees of traits, shown in \cref{maintst} for~\cite{Big5onlinedata2} and in \cref{smalltst} for~\cite{Big5onlinedata1}, display a hier\-archy between those traits. This has two aspects. The first aspect is the structure of the tree: this tells us how the traits are related to each other, i.e., which of them refines which. The second aspect is at which level the various traits occur: this tells us their complexity, cohesion, and visibility (\cref{visibility}).

\goodbreak

Let us look at the first of these aspects. Traits sitting above other traits in the figure can be viewed as a common generalisation of those. The trait~$T_8$ in \cref{maintst}, for example, is shown above the traits $A$ and~$E$, so it is something like a common generalisation of the two. Indeed, the partitions in~$S$ oriented by the tangles in~$T_8$ are oriented in the same way by the tangles in~$A$ and by the tangles in~$E$. Those latter tangles, however, may disagree on their orientation of partitions whose order is too large for them to be oriented by the tangles in~$T_8$. In particular, they disagree on their unique efficient distinguisher, the partition~$s$ of lowest order which they orient differently. Since $A$ and~$E$ both refine~$T_8$, the order of~$s$ is at least that of the largest tangle in~$T_8$.

By contrast, the trait~$T_2$ is an even broader common generalisation of $A$ and~$E$, one that also encompasses~$O$~-- but neither $C$ nor~$N$. The traits $A$ and~$E$ are distinguished from~$O$ by the same efficient distinguisher~$s$, whose order is at least that of the largest tangle in~$T_2$. This~$s$ is also the efficient distinguisher of $T_8$ and~$O$, but it does not distinguish $A$ from~$E$.

Let us now turn to the second aspect of \cref{maintst}, the levels at which the various traits are shown. The vertical scale in the figure indicates order: of the tangles representing the traits depicted, and of the partitions of~$Q$ in~$S$ that distinguish them.%
   \footnote{But note that \cref{maintst} is not drawn to scale. The lengths of the lines, which we shall see indicate the visibility of traits, thus have to be compared with caution: lines further up span a smaller vertical difference than their lengths appear to indicate.}

The trait~$T_6$, for example, is entered in the diagram at level~59. This means that the largest order of a tangle in this trait is~59. The lowest order of a tangle in~$T_6$ can also be read out of the picture: it is the lowest order at which the highest-order tangle in the parent of~$T_6$, trait~$O$, splits into tangles belonging to its two children, the traits~$T_4$ and~$T_6$. As $O$ is entered at level~43, this order of the efficient distinguisher between the tangles in~$T_4$ and those in~$T_6$ is at least~43.%
   \footnote{More precisely, it is the lowest order of a separation~$s$ not oriented by the largest tangle in~$O$. Since that tangle has order~43, it orients precisely the separations in~$S$ that have order less than~43. So $s$ is the element of~$S$ of the smallest order that is at least~43.}
   If it is exactly~43, the trait~$T_6$ has complexity~43, cohesion~59, and visibility $59-43+1=17$. Its visibility is indicated roughly%
   \COMMENT{}
   by the length of the line that joins $T_6$ to~$O$ in the figure. This is higher than the visibility of the trait~$T_5$, say, but lower than that of the traits~$C$ and~$N$.%
   \COMMENT{}

A particularly interesting aspect of the vertical distribution of the traits in \cref{maintst} is that some of the $OCEAN$ traits are born only when others have already died: there is no order of separations in~$S$ at which they overlap. Traits $A$ and~$E$, for example, are born just below level~940, when their parent trait~$T_8$ splits. At this time the traits~$C$ and~$O$ are already dead. Trait~$E$ then outlasts trait~$A$ by a considerable margin, as does trait~$N$, which was born even before $A$ and~$E$.

{\sl In summary, while the `big five traits' targeted by~$Q$ do show up as tangle traits, they do so at very different times, or levels of `resolution' of the lens~-- order~-- through which we view our tangles and the traits to which they combine.}%
   \COMMENT{}

\medbreak

Let us now turn to the smaller study of~\cite{Big5onlinedata1}. Its tree of traits is depicted in \cref{smalltst}. For any comparison with the tree of traits for the larger study~\cite{Big5onlinedata2}, note that while the respective traits $O,C,E,A,N$ were named according to the same rule, applied independently to the two datasets, the enumerations $T_1,T_2,\dots$ and $T'_1,T'_2,\dots$ of the other traits are arbitrary and not related.

The most important result on which the two studies agree is that both feature traits that merit being called $O,C,E,A,N$ according to our rule in \cref{traitsfound}: their lowest-order tangle is guided by the subset of~$Q$ specified by the letter that names them, but by no other such subset, and this applies to no smaller trait.%
   \COMMENT{}

One structural similarity between~\cite{Big5onlinedata1} and~\cite{Big5onlinedata2} is that the traits $A$ and~$E$ are direct siblings: in both studies they have a common parent (the trait~$T_8$ in~\cite{Big5onlinedata1} and the trait~$T'_4$ in~\cite{Big5onlinedata1}). And in both datasets the trait~$A$ then splits into two more traits, while $E$ does not have distinct refinements.

A~closer look also reveals some interesting differences between~\cite{Big5onlinedata1} and~\cite{Big5onlinedata2}.

A~particularly striking feature of~\cite{Big5onlinedata1} is that the $OCEAN$ traits are, successively, split off one line of refining non-$OCEAN$ traits, the traits $T'_1 > T'_2 > T'_3 > T'_4$ (where $>$ indicates refinement as shown in the figure). In~\cite{Big5onlinedata2}, by contrast, the the $OCEAN$ traits form two groups, $C,N$ versus $A,E,O$, where $A$ and~$E$ form a subgroup of the latter group~$A,E,O$. These two major groups are in fact separated by {\em two\/} levels of more general traits: by the trait~$T_1$ (which the two groups refine in different ways), but also by its two children, the traits $T_2$ and~$T_3$. Given the fact that $T_2$ and~$T_3$ are themselves traits, this seems quite remarkable~-- and potentially interesting from an interpretation point of view. While we limit ourselves to structural comparisons here, we shall see in \cref{interpretation} that it is possible to use the information from the Appendix to write down how these traits orient their distinguishers. This provides experts in psychology with direct interpretable information.

In~\cite{Big5onlinedata2}, the traits $C$ and~$N$ have a direct common generalisation, the trait~$T_3$. In~\cite{Big5onlinedata1}, they also have a common generalisation that does not also generalise any of the other $OCEAN$ traits, the trait~$T'_2$. This, however, is not a direct generalisation just of the two traits $C$ and~$N$: there is an intermediate trait~$T'_3$, which is already a generalisation of~$N$ (though not of~$C$), and the common generalisation~$T'_2$ of~$C$ and~$N$ also generalises~$T'_3$. The study of~\cite{Big5onlinedata1} thus sees more structure than~\cite{Big5onlinedata2} in the relationship between the traits $C$ and~$N$.

Another interesting difference between~\cite{Big5onlinedata1} and~\cite{Big5onlinedata2} is that, in~\cite{Big5onlinedata1}, the trait~$O$ does not further refine, while in~\cite{Big5onlinedata2} it splits into traits~$T_6$ and~$T_4$, which in turn splits into $T_5$ and~$T_7$. Thus, \cite{Big5onlinedata2}~reveals substantial structure inside the trait~$O$, which~\cite{Big5onlinedata1} did not see.

\subsection{Interpretation}\label{interpretation}

Tangles, unlike clusters found by some more traditional clustering algorithms, are in principle open to interpretation: every tangle is a set of oriented partitions of~$Q$, and these are known explicitly. An expert psychologist can draw interpretative conclusions from the way our tangle traits orient these partitions.

In our particular case, we can even say something interpretative about the tangle traits we found without any expert knowledge of psychology. This is because we were able to relate some of these traits, $O$, $C$, $E$, $A$ and~$N$, to five groups of questions in~$Q$, as indicated in \cref{traitsfound}. This already yields interpretative information for these five tangle traits themselves, if one considers these five sets of questions as understood at the interpretation level. And this is a reasonable assumption: recall that these five sets of questions were {\em designed\/} to flesh out some {\em interpretation\/} of the Five Factors, see \cref{interpretation1} and the Introduction. Our five tangle traits named $O,C,E,A,N$ therefore inherit those interpretations in the way made precise in \cref{traitsfound}.

So what about interpreting the other ten of our fifteen tangle traits in~\cite{Big5onlinedata2},~say? Our structural tangle analysis in \cref{traitstructure} yields that some of those other ten traits are common generalisations of specific subsets of the $OCEAN$ traits, and thereby inherit their interpretations. But even those traits that are not common generalisations of $OCEAN$ traits can be interpreted succinctly, as follows.

Each of the 15 traits in~\cite{Big5onlinedata2} is determined uniquely by which of the efficient tangle distinguishers it orients, and how. While there are seven efficient distinguishers in total, one for every branching node of the tree of traits (\cref{maintst}), the tangles in a given trait orient only those distinguishers whose order is smaller than their complexity. But different traits are distinguished by at least one of these distinguishers, so each trait is identified uniquely by how it orients the efficient distinguishers that it does orient.

And every oriented distinguisher is directly accessible to interpretation: it is one concrete subset of~$Q$. For the efficient trait distinguishers, these subsets are listed explicitly in the Appendix, in \cref{distinguishers}. From these lists one can combine, for each of our 15 traits~$T$ other than the all-encompassing~$T_1$, how the tangles in~$T$ orient the distinguishers~$s$ which they orient at all, those of order less than the complexity of~$T$.%
   \COMMENT{}
   Conversely, given any two traits~$T$ and~$T'$, \cref{maintst} tells us which~$s\in S$ is their unique efficient distinguisher, and its list in \cref{distinguishers} then tells us how $T$ and~$T'$ orient~$s$ towards its opposite sides and which concrete subsets of~$Q$ those sides are.

A~more direct, and simpler, route to interpreting our tangle traits is to identify, for each given trait, the questions in~$Q$ that represent it best~-- in a precise, formal sense. We discuss this next.%
   \COMMENT{}

\subsection{Which questions represent the traits best?}\label{point-tangle-sim}

Recall that our questionnaire~$Q$ was designed for the purpose of testing the five OCEAN traits, originally in the pre-formal sense of `trait' discussed and criticised in \cref{intro}. In \cref{traitsfound} we gave a formal definition of when we denote any of our tangle traits (which, in contrast, are a formal concept) by those $OCEAN$ labels. For these five traits, but also for all the other six (respectively, ten) traits we found in \cite{Big5onlinedata1} and~\cite{Big5onlinedata2}, our tangle framework allows us to determine explicitly which of the 50 questions in~$Q$ are best suited to test for that particular trait. This works as follows.

Every tangle in~$\vS$, by definition, orients some of the partitions of~$Q$ towards one of their two sides. Every element of~$Q$ does that too: we can think of it as orienting every partition of~$Q$ towards the side that contains it. We can therefore compare the questions $q$ in~$Q$ directly with the tangles~$\tau$ in~$\vS$, simply by counting the number~$\sigma(q,\tau)$ of partitions they orient in the same way (\cref{point-tangle_sim_def}).

In order to assess how well a question $q$ represents an entire trait~$T$, now, rather than a given tangle, it would be natural to simply use $\sigma(q,\tau)$ for some $\tau\in T$. But which~$\tau\in T$ should we choose for this comparison? Fortunately, it does not matter: as long as we compare $q$ and~$\tau$ only on partitions $s\in S$ that really matter, the result will be the same.

Indeed, given $q\in Q$ and a tangle trait~$T$, write $k^-$ for the complexity of~$T$ and $k^+$ for its cohesion. Then the tangles in~$T$ agree on all the partitions in
 $$S_T := \{\,s\in S : k^- \le |s|+1\le k^+\}$$
 which they orient,%
   \COMMENT{}
   and they disagree, with any tangle that can be distinguished from them at all, on all partitions in~$S_T$ that such a tangle also orients. The partitions in~$S_T$, therefore, are those on which the trait~$T$ comes into its own: those on which its tangles differ from tangles in all traits that are essentially different from~$T$.%
   \COMMENT{}

We now define $T(s) := \tau(s)$ for all $s\in S_T$, where $\tau$ is any tangle in~$T$ that orients~$s$. (We do not have to specify which~$\tau$, since they all agree on~$s$.)%
   \COMMENT{}
   Then
  $$\sigma(q,T) := \big|\{\,s\in S_T : q(s) = T(s)\,\}\big|$$
 measures what we want: the similarity between $q$ and the tangles in~$T$ on the partitions that reflect~$T$ best.

\cref{traitsrepresented} lists, for each of the traits found in~\cite{Big5onlinedata1} and~\cite{Big5onlinedata2}, the three questions $q$ in~$Q$ that represent this trait~$T$ best as measured by~$\sigma(q,T)$.%
   \footnote{The questions themselves are listed in~\cref{Q}.}
   These tables contain a wealth of information that is directly accessible to interpretation. Not only are the individual traits each represented by three concrete questions; it is also instructive to compare how the structural relationship between the traits, as expressed by the trees of traits shown in \cref{smalltst,maintst}, compares with the sets of questions that represent them. 

For example, \cref{maintst} shows that the trait~$O$ in~\cite{Big5onlinedata2} splits into the subtraits $T_4$ and~$T_6$. These are represented in \cref{traitsrepresented} by disjoint sets of questions: $T_4$~by $O_4,O_5,O_{10}$ and $T_6$ by $O_1,O_7,O_8$, which helps us distinguish them also at the interpretation level. The subtrait~$T_4$ then splits again, into $T_5$ and~$T_7$. But these are both represented best by $O_5,O_6,O_{10}$, the same set of questions: their difference as subtraits of~$T_4$ is more subtle than what their representing questions indicate. But this difference is captured succinctly by one explicitly known partition: their efficient distinguisher. This is listed in \cref{distinguishers}, associated with~$T_4$.

We must leave any further analysis to the experts in psychology, contenting ourselves with pointing out how the information on this can be based is made explicit in our tangle data.

\subsection{Robustness of our methods}\label{robustness}

Recall that by {\em robustness\/} of our methods we mean the invariance of our findings under small changes to the input, while we keep those methods unchanged. To test this, we repeated our tangle analysis of the large study~\cite{Big5onlinedata2} described earlier in this section on ten subsets of its participants.

We first divided its entire set of about a million participants into 5 disjoint subsets of 200.000 participants and analysed their tangles. We then chose five further random subsets of about 10.000 participants each, and again analysed their tangles.

\begin{figure}[htb]
    \center
	\includegraphics[scale=0.8]{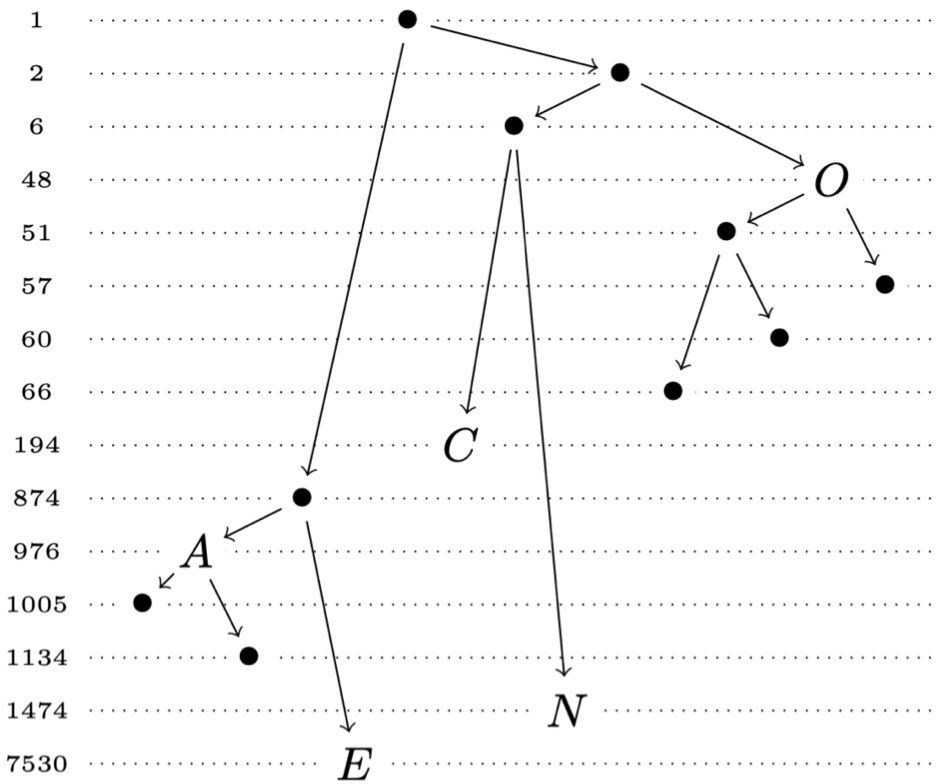}
	\caption{The only different tree of traits on a large subset of~\cite{Big5onlinedata2}}\vskip-3pt\vskip0pt
	\label{5thlargesubset}
\end{figure}

The trees of traits we found in the five large subsets were very similar to that of the entire dataset (\cref{maintst}). Two of them were in fact identical to it. Another two were identical to it except that trait~$A$ lacked the two children it had in \cref{maintst}, denoted there as $T_9$~and~$T_{10}$. The fifth tree of tangles is shown in \cref{5thlargesubset}. It differs from the other four in that the two long branches hanging off $T_2$ and~$T_3$ in \cref{maintst} are swapped. So there is now a joint trait%
   \COMMENT{}
   generalising $C,N,O$ but not $A,E$, whereas in \cref{maintst} there was one generalising $A,E,O$ but not $C,N$. The $OCEAN$ traits themselves are independent,%
   \COMMENT{}
   as before.

\begin{figure}[htb]
	\center
	\includegraphics[scale=0.58]{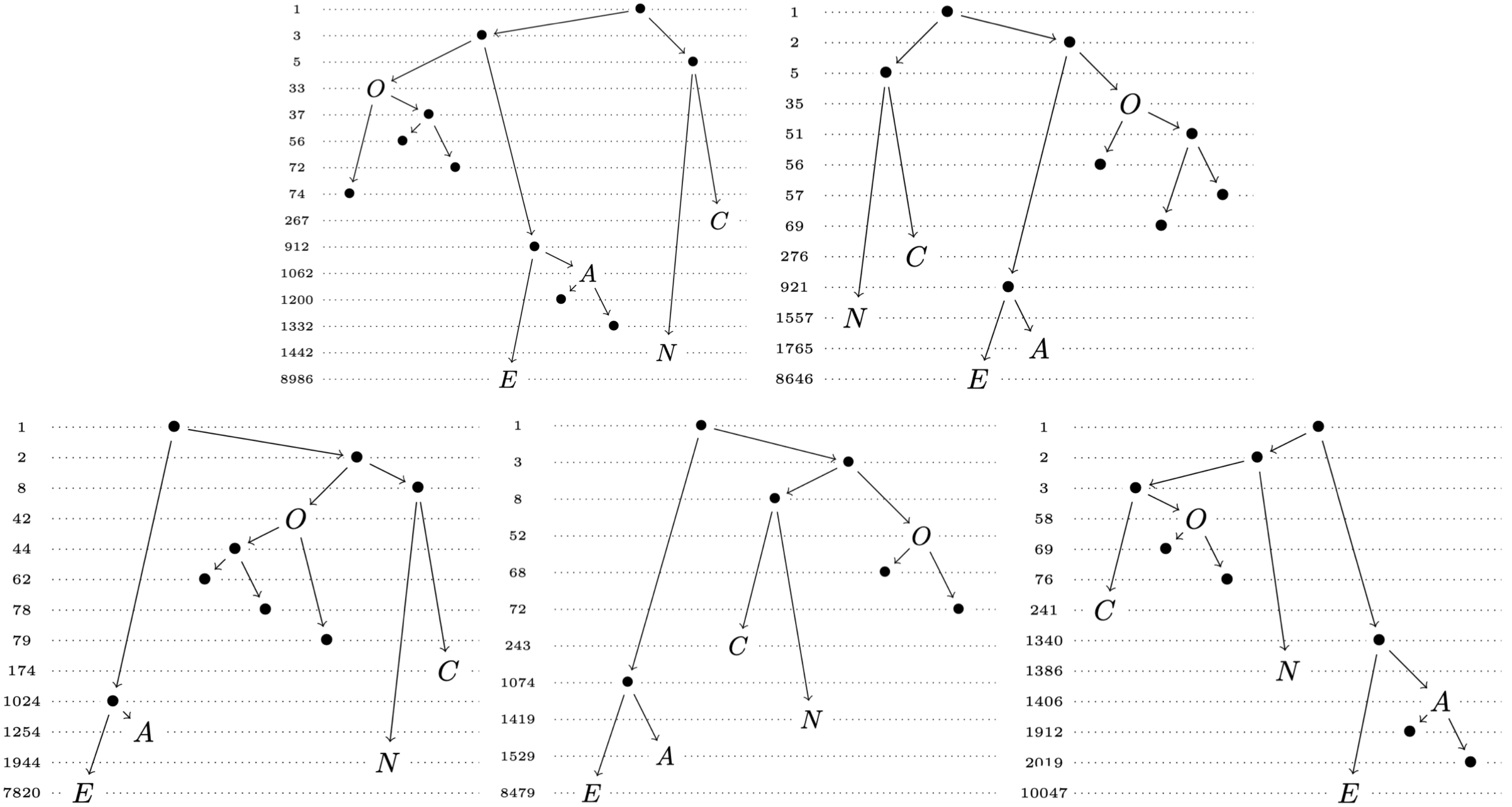}
	\caption{The trees of traits for the five small subsets of~\cite{Big5onlinedata2}}\vskip-9pt\vskip0pt
	\label{5smallsubsets}
\end{figure}

The trees of traits we found on the five smaller subsets of~\cite{Big5onlinedata2} are depicted in~\cref{5smallsubsets}. They show a little more variation~-- not unexpected for smaller datasets. The first is again identical to that of \cref{maintst}. The second is identical to it except that $A$ has no children, as earlier. The other three are variants of the tree in \cref{5thlargesubset}.%
   \footnote{Note that for the tree structure it is irrelevant which two branches emanating from a node is drawn left and which right. The branch containing $C$, $N$ and~$O$ in last tree in \cref{5smallsubsets}\COMMENT{} is in fact identical to that in the tree before\COMMENT{tst2}, except that the parent trait of $C$ and~$N$ has split into two traits.}

Thus, the traits we found in the ten subsets of~\cite{Big5onlinedata2}, as well as their tree structure, are all very similar to those for the entire dataset (\cref{maintst}). In each of them there are five traits that merit being denoted as $O,C,E,A,N$ by our rule in \cref{traitsfound}. Their relative structure is also the same as before: traits $A$ and~$E$ combine to a more comprehensive trait not generalising any other $OCEAN$ traits. And so do $C$ and~$N$, except in the last tree of \cref{5smallsubsets}. In six of the ten trees,%
   \COMMENT{}
   $O$~combines with the supertrait of $A$ and~$E$ to a trait not generalising $C$ or~$N$, as in \cref{maintst}, while in the other four%
   \COMMENT{}
   it combines with $C$ and~$N$%
   \COMMENT{}
   to a trait not generalising $A$ or~$E$.

Overall, our tangle analysis of the ten subsets of~\cite{Big5onlinedata2} yields essentially the same results as that of the entire dataset. This means that our tangle analysis is robust in the sense defined earlier: the essential aspects of its results are invariant under changes of data that are not due to other factors, such as a different time when the study was conducted, or a different sample of participants.

If we take this as confirmation of the robustness our methods also with respect to our analysis of the data from~\cite{Big5onlinedata1}, the implication is that the difference our analysis in \cref{traitstructure} found between the studies of \cite{Big5onlinedata1} and~\cite{Big5onlinedata2} reflects genuine difference in the data. This, in turn, may have interesting implications for their interpretation, since the differences between \cite{Big5onlinedata1} and~\cite{Big5onlinedata2}, as well as what they have in common (such as being based on the same questionnaire~$Q$), are known explicitly.

\subsection{Stability}\label{stability}

Recall that by {\em stability\/} of our findings we mean their invariance under small changes of our methods, performed on the same input.

In our methods we initially made some reasonable choices that we kept fixed throughout our tangle analysis. Those included the use of the ratio cut weight order function (based, however, on a similarity function which we did vary), to require our tangles to have a minimum agreement of~2, and the various normalisations described in \cref{NormalisingAnswers}.

However we tested variation of two parameters: of our choice of the similarity function used by the ratio cut weight order function (see \cref{SimilarityFunctions}), and of our choice of the matrix ($L$~or~$J$, see \cref{ChoosingPartitions}) whose eigenvectors provided the initial set of partitions of~$Q$  from which we then built our final set~$S$ of partitions.%
   \COMMENT{}\looseness=-1

We found that replacing $L$ with~$J$ made no structural difference to the traits we found and their relationships. The only difference lay in the complexity and cohesion of the traits, but with very little resulting difference for their visibilities. So our findings were perfectly stable in this respect.

This is not unremarkable, and we interpret it as an indication of reliability for the tangle method in clustering in general. Its fundamental idea is to ignore just the right amount of detail in a fuzzy environment to identify the `true' clusters and their mutual relationship. Both these turned out to be identical in our tangle analysis even when its input data, the starting set~$S$ of partitions of~$Q$, was very different as we obtained it from the eigenvectors of the matrices $L$ and~$J$, respectively.

\begin{figure}[htb]
	\center
	\includegraphics[scale=0.65]{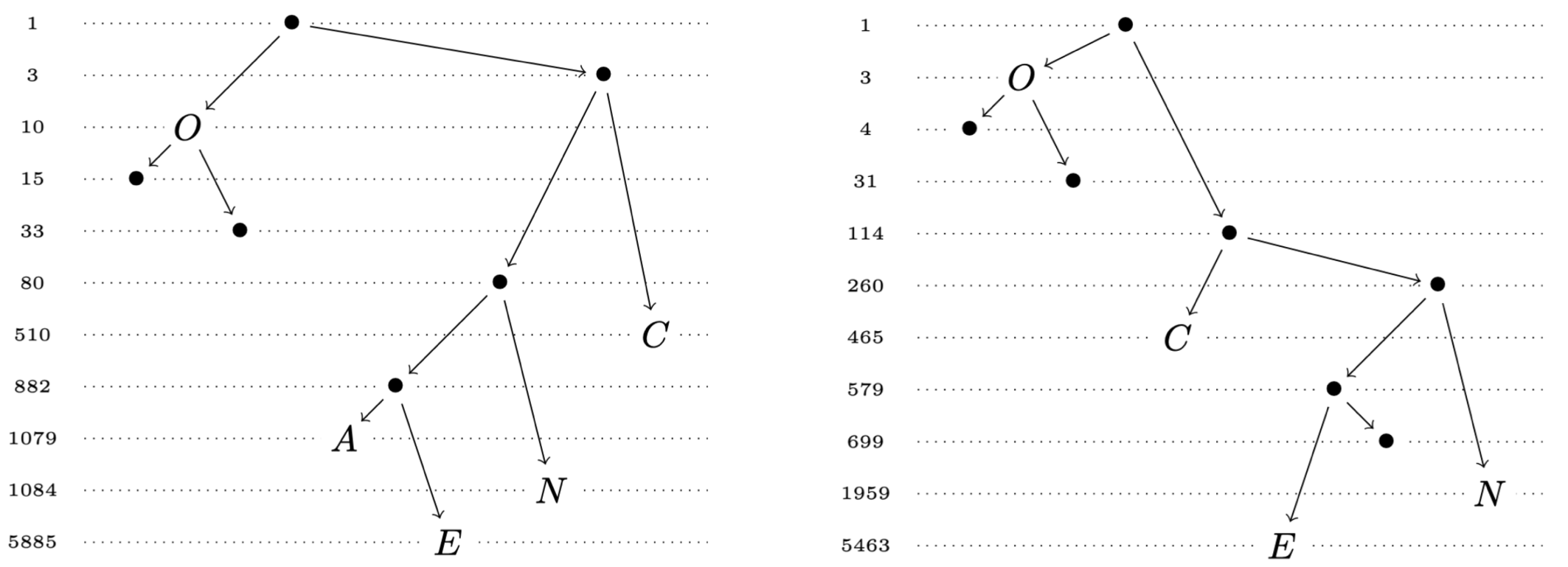}
	\caption{The trees of traits computed with cosine similarity. The tree for the large study~\cite{Big5onlinedata2} is shown on the left, that for the small study~\cite{Big5onlinedata1} on the right.}%
   \COMMENT{}
	\label{bothcosine}
\end{figure}

Let us now report briefly what we found when we based the ratio cut weight order, which we used throughout, on cosine similarity (\cref{cosimdef}) rather than on entropy similarity (\cref{entsimdef}). The trees of traits we found in \cite{Big5onlinedata1} and~\cite{Big5onlinedata2}, respectively, are shown in \cref{bothcosine}.

Once more, we used the labels of $O,C,E,A,N$ for traits that corresponded to the subsets of~$Q$ bearing these names, according to the rule from \cref{traitsfound}. This time, no trait found in the smaller study of~\cite{Big5onlinedata1} could be named as~$A$. But it had traits corresponding to $O,C,E$ and~$N$, while in the larger study of~\cite{Big5onlinedata2} all five $OCEAN$ traits occurred. 

Unlike for the traits based on entropy similarity, the trees of traits were now strikingly similar for the smaller and the larger study, \cite{Big5onlinedata1} and~\cite{Big5onlinedata2}. The only structural difference%
   \COMMENT{}
   between the two is that the sibling of~$E$ is~$A$ in~\cite{Big5onlinedata2}, while in~\cite{Big5onlinedata1} it is an unmarked trait. Both $A$ and this unmarked trait have low visibility in their respective tree. More detailed analysis shows that the unmarked trait is in fact quite close to meeting our standards for being named as~$A$, whereas trait~$A$ in~\cite{Big5onlinedata2} meets these only just.

An~explicit look at the partitions of~$Q$ we collected into~$S$ using cosine rather than entropy similarity indicates that they are rougher: the entropy-based low-order partitions conform more readily to what, from an interpretation point of view, we would expect to be relevant partitions of~$Q$. But this roughness is smoothed again by the tangles they form, and even more so by the traits to which these combine: we get a similar number of traits as before, the expected $OCEAN$ traits are among them, and there are further traits that refine or generalise these $OCEAN$ traits, thereby endowing them with additional structure.%
   \COMMENT{}\looseness=-1

Finally, there is a striking structural similarity between the two trees of traits based on cosine similarity (\cref{bothcosine}) with the tree of traits based on entropy similarity (\cref{smalltst}) for the smaller study~\cite{Big5onlinedata1}. Indeed, in structural terms the tree of traits shown on the left in \cref{bothcosine} (which is based on the {\em larger\/} study~\cite{Big5onlinedata2}) differs from that of \cref{smalltst} (for the {\em smaller\/} study~\cite{Big5onlinedata1}) only in that $O$ splits into two subtraits in the former but not in the latter, while $A$ splits into two subtraits in the latter but not in the former. Otherwise, the two trees are structurally identical.%
   \footnote{In mathematical terms: once we delete the two pairs of children named above, there exists a graph isomorphism between the two trees that fixes their $OCEAN$ traits, i.e., which maps corresponding $OCEAN$ traits to each other.}
   The similarity between these two trees seems remarkable because they are based on different studies, \cite{Big5onlinedata1} and~\cite{Big5onlinedata2}, which in our earlier analysis we found to exhibit more distinct trait structures (which, moreover, we found to be robust, see \cref{robustness}).

We cannot explain this phenomenon. The fact that the trees of traits are more similar for \cite{Big5onlinedata1} and~\cite{Big5onlinedata2} under cosine similarity (\cref{bothcosine}) than they are under entropy similarity (\cref{maintst,smalltst}) can be interpreted as an indication that tangle analysis based on cosine similarity is a blunter tool than tangle analysis based on entropy similarity.%
   \COMMENT{}
   But the fact that the more sensitive tool yields, for one of these studies~\cite{Big5onlinedata1}, results very similar to those returned by the blunter tool for both studies, while it returns reliably different results for the other study~\cite{Big5onlinedata2}, is harder to interpret. It might be seen as an indication of an anomaly in the design of~\cite{Big5onlinedata2}, which analysis based on cosine similarity is (perhaps beneficially) too blunt to pick up, or it may be an artefact that indicates nothing significant.\looseness=-1

\section{Conclusion}

The problem this paper aimed to address was to bridge the gap between the mathematical notion of a {\em factor\/}, in the factor analysis performed on personality data from Cattell~\cite{Cattell45} and Fiske~\cite{Fiske49} by Tupes and Christal~\cite{Big5Origin}, and interpretations of these (`big five') factors on which questionnaires for personality assessment are currently based. The gap stems from the fact that factors, by definition, are abstract linear combinations of answers obtained for the studies analysed, while questionnaires testing for such factors are formulated based not on those linear combinations of the original questions but on new questions designed to test for an {\em interpretation\/} of those factors, conceived of as a `trait'.

To do this, we developed quantitative criteria for groups of questions to merit being thought of as testing for any common trait at all. This required us to provide a formal definition of `trait', against which such questionnaires could then be tested.

We first proposed, in general clustering terms, two criteria for a group of questions to achieve this: as `extensional cohesion' and `extensional completeness'. Our use of the novel clustering method of {\em tangle analysis\/} then enabled us to propose a rigorous notion of `trait', on which we based our further investigations: that of an equivalence class of tangles under some generic equivalence relation from tangle theory, whose origin lies in the tangle analysis of visual~data. This formal notion of (tangle) trait satisfies our earlier requirements of extensional cohesion and completion; indeed these are built in to the very notion of a tangle and of tangle equivalence.%
   \COMMENT{}
   But it goes further, in that it identifies the structural essence of clusters in a fuzzy environment, while filtering out unnecessarily detailed but less essential data as `noise'.

We analysed the data returned by the separate studies of \cite{Big5onlinedata1} and~\cite{Big5onlinedata2}, both for the same questionnaire~$Q$ designed to test for the `big five' OCEAN traits. Our analysis confirms that the five groups of questions in~$Q$ designed to test for those five `traits' do indeed each test for something that satisfies our formal notion of a tangle trait, with some reservations in the case of trait~$A$ (`agreeableness'). But in addition to the five $OCEAN$ traits we also found some others, both common generalisations and refinements of $OCEAN$ traits: six in the smaller study of~\cite{Big5onlinedata1}, and ten in the larger study of~\cite{Big5onlinedata2}.

These traits, including the five $OCEAN$ traits, are not equally cohesive and complete, and our analysis quantifies to what extent they are.%
   \COMMENT{}
   {\em They are also interrelated in ways that were not, to the best of our knowledge, known before.\/} These relationships are exhibited by {\em trees of traits\/} (\cref{traitstructure}). 

For each of the five $OCEAN$ traits, but also for the other traits we found, our analysis can identify the questions in~$Q$ that represent this trait best, in a rigorous, quantitative sense. In \cref{traitsrepresented} we list the top three questions for each trait. This is one aspect in which the tangle traits we found are directly interpretable~-- unlike the factors found in larger studies by factor analysis.\looseness=-1

Further interpretability of the traits we found can be read off explicitly from their {\it efficient distinguishers\/}: five in~\cite{Big5onlinedata1}, and seven in~\cite{Big5onlinedata2}. Each of these is one concrete partition of~$Q$ into two sets of questions. These are listed explicitly in \cref{distinguishers}, and are thus directly accessible to expert interpretation.

Finally, we tested the robustness of our methods (\cref{robustness}) for the larger study of~\cite{Big5onlinedata2}, as well as the stability of our findings under changes to our tangle parameters (\cref{stability}). We found that both the traits found and their relationships as expressed in the trees of traits were essentially invariant under changes of input data taken disjointly and independently from the same study. We found that our results were perfectly stable under changes to the parameters $L$ or~$J$ that determined the partitions of~$Q$ for our tangles, and largely stable when we replaced the entropy-based similarity measure we used with the cruder measure of cosine similarity. Both these are state-of-the-art similarity measures in information theory and data science.

In summary, we found that the method of tangle-based clustering can be used reliably to find previously unknown traits in existing data obtained for FFM-based personality tests, and to relate known traits to each other, all in a rigorously defined quantitative sense.

\bibliographystyle{abbrv}
\bibliography{collective}

\section{Appendix}

\subsection{The questionnaire~\boldmath $Q$}\label{Q}

Here is a list of the 50 questions in the questionnaire~$Q$ used in \cite{Big5onlinedata1} and~\cite{Big5onlinedata2}:

\medskip\noindent{\bf
Openness}
\begin{itemize}
    \item[$O_1$]	I have a rich vocabulary.
\item[$O_2$]	I have difficulty understanding abstract ideas.
\item[$O_3$]	I have a vivid imagination.
\item[$O_4$]	I am not interested in abstract ideas.
\item[$O_5$]	I have excellent ideas.
\item[$O_6$]	I do not have a good imagination.
\item[$O_7$]	I am quick to understand things.
\item[$O_8$]	I use difficult words.
\item[$O_9$]	I spend time reflecting on things.
\item[$O_{10}$]	I am full of ideas.
\end{itemize}

\medskip\noindent{\bf
Conscientiousness}
\begin{itemize}
 \item[$C_1$]	I am always prepared.
\item[$C_2$]	I leave my belongings around.
\item[$C_3$]	I pay attention to details.
\item[$C_4$]	I make a mess of things.
\item[$C_5$]	I get chores done right away.
\item[$C_6$]	I often forget to put things back in their proper place.
\item[$C_7$]	I like order.
\item[$C_8$]	I shirk my duties.
\item[$C_9$]	I follow a schedule.
\item[$C_{10}$]	I am exacting in my work.
\end{itemize}

\medskip\noindent{\bf
Extraversion}
\begin{itemize}
    \item[$E_1$]	I am the life of the party.
    \item[$E_2$]	I don't talk a lot.
    \item[$E_3$]	I feel comfortable around people.
    \item[$E_4$]	I keep in the background.
    \item[$E_5$]	I start conversations.
    \item[$E_6$]	I have little to say.
    \item[$E_7$]	I talk to a lot of different people at parties.
    \item[$E_8$]	I don't like to draw attention to myself.
    \item[$E_9$]	I don't mind being the center of     attention.
    \item[$E_{10}$]	I am quiet around strangers.
\end{itemize}

\medskip\noindent{\bf
Agreeableness}
\begin{itemize}
    \item[$A_1$]	I feel little concern for others.
\item[$A_2$]	I am interested in people.
\item[$A_3$]	I insult people.
\item[$A_4$]	I sympathize with others' feelings.
\item[$A_5$]	I am not interested in other people's problems.
\item[$A_6$]	I have a soft heart.
\item[$A_7$]	I am not really interested in others.
\item[$A_8$]	I take time out for others.
\item[$A_9$]	I feel others' emotions.
\item[$A_{10}$]	I make people feel at ease.
\end{itemize}

\medskip\noindent{\bf
Neuroticism}
\begin{itemize}
    \item[$N_1$]	I get stressed out easily.
\item[$N_2$]	I am relaxed most of the time.
\item[$N_3$]	I worry about things.
\item[$N_4$]	I seldom feel blue.
\item[$N_5$]	I am easily disturbed.
\item[$N_6$]	I get upset easily.
\item[$N_7$]	I change my mood a lot.
\item[$N_8$]	I have frequent mood swings.
\item[$N_9$]	I get irritated easily.
\item[$N_{10}$]	I often feel blue.
\end{itemize}

\subsection{All traits, represented by typical questions}\label{traitsrepresented}

In \cref{point-tangle-sim} we saw that tangles in~$\vS$, and by implication also tangle traits~$T$, can be compared directly with individual questions~$q$ from~$Q$ (see~\cref{Q}), by counting how many of the partitions in~$S$ they orient in the same direction. We said that when this number is high the question~$q$ {\em represents\/} the trait~$T$~well.

The following two tables list the top three representatives from~$Q$, in order (best first), for each of the traits we found in \cite{Big5onlinedata1} and~\cite{Big5onlinedata2}:

{\small
\bigbreak\noindent
\begin{tabular}{l l l l l}
 {\boldmath $O$}: $O_{10},O_6,O_1$ & {\boldmath $C$}: $C_5,C_6,C_1$ & 
 {\boldmath $E$}: $E_7,E_5,E_2$  &{\boldmath $A$}: $A_9,A_1,A_8$ &
 {\boldmath $N$}: $N_6,N_8,N_9$   \\ 
 {\boldmath $T_2$}: $O_{10},C_5,N_3$ & {\boldmath $T_3$}: $C_5,N_9,N_8$ &
 {\boldmath $T_4$}: $O_{10},O_5,O_{4}$ & {\boldmath $T_5$}: $O_{10},O_6,O_{5}$ &
 {\boldmath $T_6$}: $O_8,O_1,O_7$ \\ 
 {\boldmath $T_7$}: $O_{10},O_5,O_{6}$ & {\boldmath $T_8$}: $A_5,A_7,E_2$ &
 {\boldmath $T_9$}: $A_5,A_7,A_4$ & {\boldmath $T_{10}$}: $A_2,A_7,A_5$
\end{tabular}
\vskip3pt\centerline{\small The questions best representing the traits in~\cite{Big5onlinedata2}}
\bigbreak\smallskip
\noindent
\begin{tabular}{l l l l l}
 {\boldmath $O$}: $O_2,O_1,O_8$ & {\boldmath $C$}: $C_5,C_6,C_9$ & 
 {\boldmath $E$}: $E_7,E_5,E_1$  &{\boldmath $A$}: $A_5,A_2,A_4$ &
 {\boldmath $N$}: $N_6,N_9,N_8$   \\ 
 {\boldmath $T'_2$}: $A_5,A_6,A_2$ & {\boldmath $T'_3$}: $O_{10},C_5,N_3$ &
 {\boldmath $T'_4$}: $E_1,E_2,E_3$ & {\boldmath $T'_5$}: $A_5,A_4,A_8$ &
 {\boldmath $T'_6$}: $A_2,A_7,A_5$
\end{tabular}
\vskip3pt\centerline{\small The questions best representing the traits in~\cite{Big5onlinedata1}}
\bigbreak
}

\subsection{All traits, represented by the tangle-distinguishers}\label{distinguishers}

In \cref{interpretation} we noted that traits are described most succinctly by the small collection of partitions in~$S$ that distinguish them efficiently from other traits. In this appendix we list these distinguishers explicitly, and then list for every trait how it orients them.

The efficient distinguisher~$s$ of two traits $T_1$ and~$T_2$ is naturally associated with their unique common ancestor~$T$ in the tree of traits that displays them. Indeed, by definition $s$~is the partition of~$Q$ in~$S$ of lowest order that $T_1$ and~$T_2$ orient differently. So they orient all partitions of lower order than~$|s|$ in the same way. The set of these oriented partitions is the largest-order tangle in~$T$, and $T$ has cohesion~$|s|$.

In the tables below we denote this partition~$s$ of~$Q$ as~$s(T)$. Note that $s$ is the efficient distinguisher not only for $T_1$ and~$T_2$, but for all pairs of distinct traits that both refine~$T$ but no tangle of higher cohesion. The two sides of a partition $s=s(T)$ are stated, as subsets of~$Q$, in the left and right half of the box below~$s$.

First, the efficient distinguishers from~\cite{Big5onlinedata2}, one associated with each branching node of the tree of traits in~\cref{maintst}:

\bigbreak
\begingroup\small
    \begin{tabular}{|l|r|}
    \hline
    \multicolumn{2}{|c|}{$s(T_{1})$} \\
    \hline
     & E1 E2 E3 E4 E5 E6 E7 E8 E9 E10  \\N1 N2 N3 N4 N5 N6 N7 N8 N9 N10  &  \\A3  & A1 A2 A4 A5 A6 A7 A8 A9 A10  \\C1 C2 C3 C4 C5 C6 C7 C8 C9 C10  &  \\ & O1 O2 O3 O4 O5 O6 O7 O8 O9 O10  \\
    \hline
    \end{tabular} \\[4pt]
    
    \begin{tabular}{|l|r|}
    \hline
    \multicolumn{2}{|c|}{$s(T_{2})$} \\
    \hline
    E1 E2 E3 E4 E5 E6 E7 E8 E9 E10  &  \\ & N1 N2 N3 N4 N5 N6 N7 N8 N9 N10  \\A1 A2 A3 A4 A5 A6 A7 A8 A9 A10  &  \\ & C1 C2 C3 C4 C5 C6 C7 C8 C9 C10  \\ & O1 O2 O3 O4 O5 O6 O7 O8 O9 O10  \\
    \hline
    \end{tabular} \\[4pt]
    
    \begin{tabular}{|l|r|}
    \hline
    \multicolumn{2}{|c|}{$s(T_{3})$} \\
    \hline
     & E1 E2 E3 E4 E5 E6 E7 E8 E9 E10  \\ & N1 N2 N3 N4 N5 N6 N7 N8 N9 N10  \\ & A1 A2 A3 A4 A5 A6 A7 A8 A9 A10  \\C1 C2 C3 C4 C5 C6 C7 C8 C9 C10  &  \\O1 O2 O3 O4 O5 O6 O7 O8 O9 O10  &  \\
    \hline
    \end{tabular} \\[4pt]
    
    \begin{tabular}{|l|r|}
    \hline
    \multicolumn{2}{|c|}{$s(O)$} \\
    \hline
     & E1 E2 E3 E4 E5 E6 E7 E8 E9 E10  \\N1 N2 N3 N4 N5 N6 N7 N8 N9 N10  &  \\A3  & A1 A2 A4 A5 A6 A7 A8 A9 A10  \\C1 C2 C3 C4 C5 C6 C7 C8 C9 C10  &  \\O1 O8  & O2 O3 O4 O5 O6 O7 O9 O10  \\
    \hline
    \end{tabular} \\[4pt]
    
    \begin{tabular}{|l|r|}
    \hline
    \multicolumn{2}{|c|}{$s(T_{4})$} \\
    \hline
     & E1 E2 E3 E4 E5 E6 E7 E8 E9 E10  \\N1 N2 N3 N4 N5 N6 N7 N8 N9 N10  &  \\ & A1 A2 A3 A4 A5 A6 A7 A8 A9 A10  \\C1 C2 C3 C4 C5 C6 C7 C8 C9 C10  &  \\O1 O2 O4 O7 O8 O9  & O3 O5 O6 O10  \\
    \hline
    \end{tabular} \\[4pt]
    
    \begin{tabular}{|l|r|}
    \hline
    \multicolumn{2}{|c|}{$s(T_{8})$} \\
    \hline
     & E1 E2 E3 E4 E5 E6 E7 E8 E9 E10  \\N1 N2 N3 N4 N5 N6 N7 N8 N9 N10  &  \\A1 A2 A3 A4 A5 A6 A7 A8 A9 A10  &  \\C1 C2 C3 C4 C5 C6 C7 C8 C9 C10  &  \\ & O1 O2 O3 O4 O5 O6 O7 O8 O9 O10  \\
    \hline
    \end{tabular} \\[4pt]
    
    \begin{tabular}{|l|r|}
    \hline
    \multicolumn{2}{|c|}{$s(A)$} \\
    \hline
     & E1 E2 E3 E4 E5 E6 E7 E8 E9 E10  \\N1 N2 N3 N4 N5 N6 N7 N8 N9 N10  &  \\A1 A3 A4 A5 A6 A8 A9  & A2 A7 A10  \\C1 C2 C3 C4 C5 C6 C7 C8 C9 C10  &  \\ & O1 O2 O3 O4 O5 O6 O7 O8 O9 O10  \\
    \hline
    \end{tabular}
   \endgroup

\bigbreak
The 14 traits in~\cite{Big5onlinedata2} other than~$T_1$ orient these distinguishers as follows. Left and right arrows correspond to the sides of~$s(T)$ shown left and right above.\medskip

\begingroup\parskip=1.8pt

 $T_{2}$: $\vs(T_{1})$ 
 
 $T_{3}$: $\sv(T_{1})$ $\vs(T_{2})$ 
 
 $O$: $\vs(T_{1})$ $\vs(T_{2})$ $\sv(T_{3})$ 
 
 $T_{4}$: $\vs(T_{1})$ $\vs(T_{2})$ $\vs(O)$ $\sv(T_{3})$ 
 
 $T_{5}$: $\vs(T_{1})$ $\vs(T_{2})$ $\vs(O)$ $\sv(T_{4})$ $\sv(T_{3})$ 
 
 $T_{6}$: $\vs(T_{1})$ $\vs(T_{2})$ $\sv(O)$ $\sv(T_{4})$ $\sv(T_{3})$ 
 
 $T_{7}$: $\vs(T_{1})$ $\vs(T_{2})$ $\vs(O)$ $\vs(T_{4})$ $\sv(T_{3})$ 
 
 $C$: $\sv(T_{1})$ $\vs(T_{2})$ $\sv(O)$ $\sv(T_{4})$ $\sv(T_{3})$ 
 
 $T_{8}$: $\vs(T_{1})$ $\sv(T_{2})$ $\vs(O)$ $\vs(T_{4})$ $\vs(T_{3})$ 
 
 $A$: $\vs(T_{1})$ $\sv(T_{2})$ $\vs(O)$ $\vs(T_{4})$ $\sv(T_{8})$ $\vs(T_{3})$ 
 
 $T_{9}$: $\vs(T_{1})$ $\sv(T_{2})$ $\vs(O)$ $\vs(T_{4})$ $\sv(T_{8})$ $\sv(A)$ $\vs(T_{3})$ 
 
 $T_{10}$: $\vs(T_{1})$ $\sv(T_{2})$ $\vs(O)$ $\vs(T_{4})$ $\sv(T_{8})$ $\vs(A)$ $\vs(T_{3})$ 
 
 $N$: $\sv(T_{1})$ $\vs(T_{2})$ $\sv(O)$ $\sv(T_{4})$ $\sv(T_{8})$ $\sv(A)$ $\vs(T_{3})$ 
 
 $E$: $\vs(T_{1})$ $\sv(T_{2})$ $\vs(O)$ $\vs(T_{4})$ $\vs(T_{8})$ $\vs(A)$ $\vs(T_{3})$ 
 
\endgroup\bigbreak

Next, the efficient distinguishers from~\cite{Big5onlinedata1}, this time associated with the branching nodes of the tree of traits in~\cref{smalltst}:

\bigbreak
\begingroup\small

    \begin{tabular}{|l|r|}
    \hline
    \multicolumn{2}{|c|}{$s(T'_{1})$} \\
    \hline
     & E1 E2 E3 E4 E5 E6 E7 E8 E9 E10  \\ & N1 N2 N3 N4 N5 N6 N7 N8 N9 N10  \\ & A1 A2 A3 A4 A5 A6 A7 A8 A9 A10  \\ & C1 C2 C3 C4 C5 C6 C7 C8 C9 C10  \\O1 O2 O3 O4 O5 O6 O7 O8 O9 O10  &  \\
    \hline
    \end{tabular} \\[4pt]
    
    \begin{tabular}{|l|r|}
    \hline
    \multicolumn{2}{|c|}{$s(T'_{2})$} \\
    \hline
    E1 E2 E3 E4 E5 E6 E7 E8 E9 E10  &  \\N1 N2 N3 N4 N5 N6 N7 N8 N9 N10  &  \\A1 A2 A3 A4 A5 A6 A7 A8 A9 A10  &  \\ & C1 C2 C3 C4 C5 C6 C7 C8 C9 C10  \\ & O1 O2 O3 O4 O5 O6 O7 O8 O9 O10  \\
    \hline
    \end{tabular} \\[4pt]
    
    \begin{tabular}{|l|r|}
    \hline
    \multicolumn{2}{|c|}{$s(T'_{3})$} \\
    \hline
     & E1 E2 E3 E4 E5 E6 E7 E8 E9 E10  \\N1 N2 N3 N4 N5 N6 N7 N8 N9 N10  &  \\ & A1 A2 A3 A4 A5 A6 A7 A8 A9 A10  \\C1 C2 C3 C4 C5 C6 C7 C8 C9 C10  &  \\O1 O2 O3 O4 O5 O6 O7 O8 O9 O10  &  \\
    \hline
    \end{tabular} \\[4pt]
    
    \begin{tabular}{|l|r|}
    \hline
    \multicolumn{2}{|c|}{$s(T'_{4})$} \\
    \hline
    E1 E2 E3 E4 E5 E6 E7 E8 E9 E10  &  \\N1 N2 N3 N4 N5 N6 N7 N8 N9 N10  &  \\A10  & A1 A2 A3 A4 A5 A6 A7 A8 A9  \\C1 C2 C3 C4 C5 C6 C7 C8 C9 C10  &  \\ & O1 O2 O3 O4 O5 O6 O7 O8 O9 O10  \\
    \hline
    \end{tabular} \\[4pt]
    
    \begin{tabular}{|l|r|}
    \hline
    \multicolumn{2}{|c|}{$s(A)$} \\
    \hline
    E1 E2 E3 E4 E5 E6 E7 E8 E9 E10  &  \\N1 N2 N3 N4 N5 N6 N7 N8 N9 N10  &  \\A2 A7 A10  & A1 A3 A4 A5 A6 A8 A9  \\C1 C2 C3 C4 C5 C6 C7 C8 C9 C10  &  \\ & O1 O2 O3 O4 O5 O6 O7 O8 O9 O10  \\
    \hline
    \end{tabular}

\endgroup

\bigbreak
The 10 traits in~\cite{Big5onlinedata1} other than~$T_1$ orient these distinguishers as follows. Left and right arrows correspond to the sides of~$s(T)$ shown left and right above.\medskip

\begingroup\parskip=\smallskipamount

 $T'_{2}$: $\vs(T'_{1})$ 
 
 $T'_{3}$: $\vs(T'_{1})$ $\sv(T'_{2})$ 
 
 $C$: $\vs(T'_{1})$ $\vs(T'_{2})$ $\sv(T'_{3})$ 
 
  $O$: $\sv(T'_{1})$ $\vs(T'_{2})$ $\sv(T'_{3})$ 

$T'_{4}$: $\vs(T'_{1})$ $\sv(T'_{2})$ $\vs(T'_{3})$ 
 
 $A$: $\vs(T'_{1})$ $\sv(T'_{2})$ $\vs(T'_{3})$ $\vs(T'_{4})$ 
 
 $T'_{5}$: $\vs(T'_{1})$ $\sv(T'_{2})$ $\vs(T'_{3})$ $\vs(T'_{4})$ $\vs(A)$ 
 
 $T'_{6}$: $\vs(T'_{1})$ $\sv(T'_{2})$ $\vs(T'_{3})$ $\vs(T'_{4})$ $\sv(A)$ 
 
 $N$: $\vs(T'_{1})$ $\sv(T'_{2})$ $\sv(T'_{3})$ $\sv(T'_{4})$ $\sv(A)$ 
 
 $E$: $\vs(T'_{1})$ $\sv(T'_{2})$ $\vs(T'_{3})$ $\sv(T'_{4})$ $\sv(A)$ 
 
\endgroup

\end{document}